\documentclass[structabstract]{aa}  
\usepackage{graphicx}
\usepackage{longtable}
\usepackage{txfonts}
\begin{document}  
\title{The supergiant B[e] star LHA 115-S 18 - binary and/or  luminous 
blue variable?\thanks{This work is partly based on observations collected 
at the European Southern Observatory (programme ID ESO 088.D-0352(A).} }
\author{J.~S.~Clark\inst{1}
\and E.~S.~Bartlett\inst{2}
\and M.~J.~Coe\inst{2}
\and R.~Dorda\inst{3}
\and F.~Haberl\inst{4}
\and J.~B.~Lamb\inst{5}
\and I.~Negueruela\inst{3}
\and A.~Udalski\inst{6}}
\institute{
$^1$Department of Physics and Astronomy, The Open 
University, Walton Hall, Milton Keynes, MK7 6AA, United Kingdom\\
$^2$The Faculty of Physical and Applied Sciences, University of Southampton, Highfield, Southampton, S017 1BJ, United Kingdom\\  
$^3$Departamento de F\'{i}sica, Ingeier\'{i}a de Sistemas y Teor\'{i}a de la Se\~{n}al, Universidad 
de Alicante, Apdo. 99,
E03080 Alicante, Spain\\
$^4$Max-Planck-Institut f\"{u}r extraterrestrische Physik, Postfach 1312,
Giessenbachstr., 85741 Garching, Germany\\
$^5$Department of Astronomy, University of Michigan, 830 Dennison 
Building, Ann
Arbor, MI 48109-1042, USA\\
$^6$Warsaw University Observatory, Aleje Ujazdowskie 4, 00-478 Warsaw, 
Poland
}

   \abstract{The mechanism by which supergiant (sg)B[e] stars support cool, 
dense dusty discs/tori and their physical relationship with other evolved, massive stars such as luminous blue variables is uncertain.}
{In order to investigate both issues  we have analysed the long term behaviour 
of the canonical sgB[e] star LHA 115-S 18.}
{We employed the OGLE II-IV lightcurve to search for (a-)periodic
variability and supplemented these data with new and historic spectroscopy.}  
{In contrast to historical expectations for sgB[e] stars, S18 is both
photometrically and spectroscopically highly variable. The lightcurve is characterised
by rapid  aperiodic `flaring' throughout the 16~years of observations.  Changes in the high excitation emission 
line component of the spectrum imply evolution  in the stellar temperature -
as expected for  luminous blue variables - although somewhat surprisingly, spectroscopic
and photometric variability appears not to be correlated. Characterised by 
emission in low excitation metallic species, the cool circumstellar torus 
appears largely unaffected by this behaviour. Finally, in conjunction with intense, highly variable
He\,{\sc ii} emission, X-ray emission implies the presence of an unseen binary companion.}
{ S18 provides observational support for the putative physical association of (a subset of) sgB[e] 
stars  and luminous blue variables. Given the nature of the circumstellar environment of S18 and that luminous blue variables  have been 
suggested as SN progenitors,  it is tempting to draw a parallel to the progenitors of SN1987A and the unusual transient SN2009ip. Moreover
the likely binary nature of S18 strengthens   the possibility that the dusty  discs/tori that characterise sgB[e] stars
 are  the result of binary-driven mass-loss; consequently  such stars may
 provide a window on the short lived phase of mass-transfer in massive compact binaries.}
\keywords{stars:evolution - stars:early type - stars:binary - star:individual:LHA 115-S 18 - 
star:emission-line}

\maketitle

\section{Introduction}

\begin{figure*}
\includegraphics[width=14cm,angle=270]{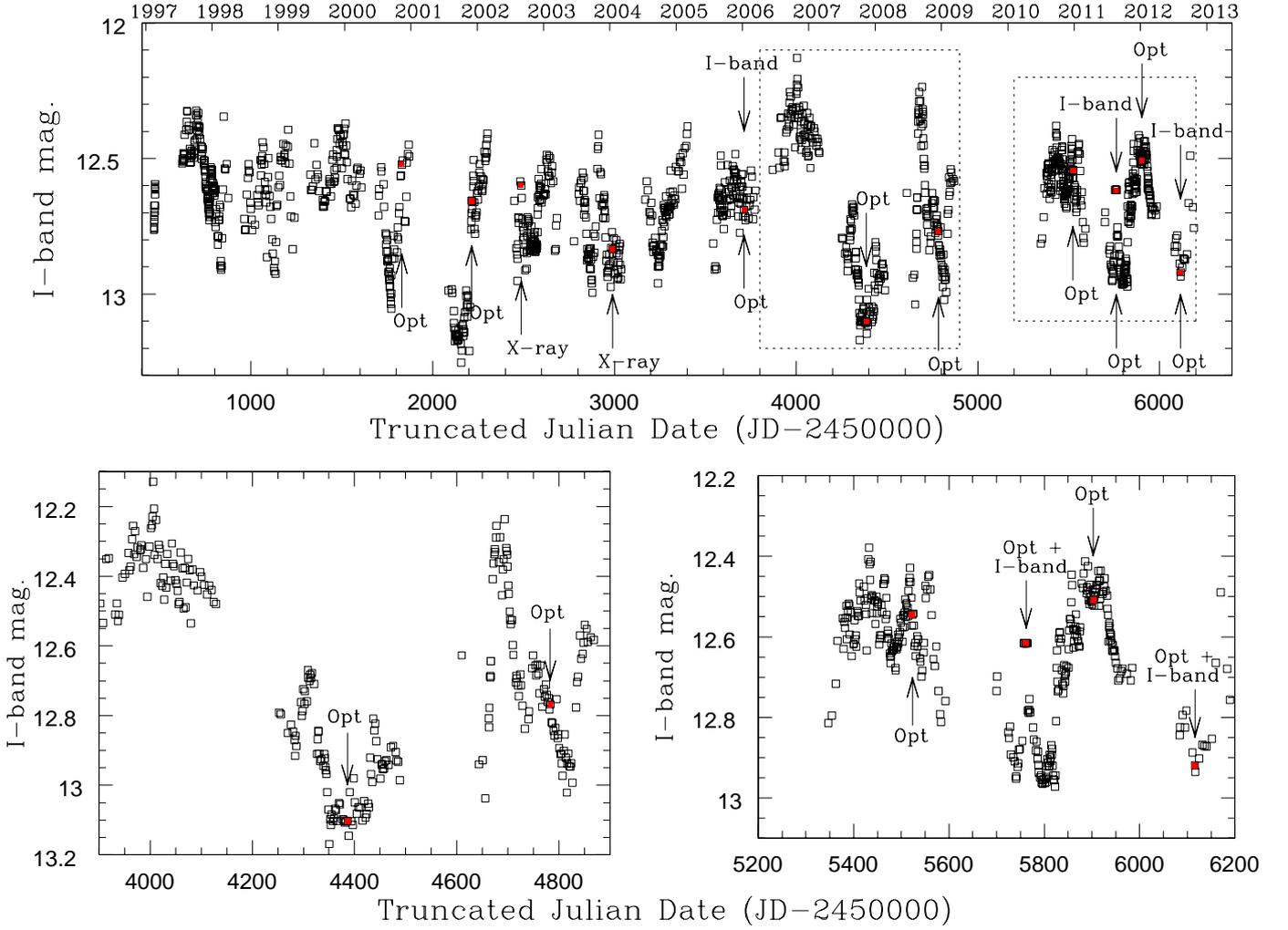}
\caption{Top panel: 
plot of the 1997-2012 OGLE-II, -III and -IV lightcurve data for LHA 115-S 18 with the timings
of optical blue- (Massey \& Duffy \cite{massey}, Torres et al. \cite{torres} and this work) and far-red (Aret et al. 
\cite{aret} and this work) spectroscopic
and X-ray (Antoniou et al. \cite{antoniou}, Bartlett et al. in prep.) observations indicated. The nearest photometric 
data points to each observation are given by the filled red symbols. Note that the spectrum  of Massey \& Duffy and 
the first epoch of Torres et al. were obtained quasi-simultaneously and the date indicated for the
first X-ray observation is the midpoint of the period over which observations were made. Photometric errors are 
 smaller than the symbol size plotted. 
 Lower panels: expanded presentation of two sub-sections of the lightcurve (indicated by dotted lines in the upper panel).}
\end{figure*}

Supergiant (sg)B[e] stars are thought to represent a short lived phase in the evolution of massive stars from 
the main sequence through to H-depleted Wolf-Rayets (WR). Observationally, they  are characterised by an 
hybrid spectrum  consisting of broad emission lines in high excitation species and narrow, low excitation 
metallic lines and a near-mid IR continuum excess. Zickgraf et al. (\cite{zickgraf85}) proposed a two 
component circumstellar environment to reconcile these findings, comprising a hot, high velocity polar 
wind and a low velocity equatorial disc or torus. The latter is expected to be either slowly expanding or 
in quasi- Keplerian rotation (Zickgraf et al \cite{zickgraf03}, Kraus et al. \cite{kraus10}, Aret et al. 
\cite{aret}) and appears to be cool and dense enough to permit both neutral gas (Zickgraf et al. 
\cite{zickgraf89}, Kraus et al. \cite{kraus07}) and dust (Kastner et al. \cite{kastner06}) to exist. 

While this  model is successful in explaining  the observations, there is currently no consensus as 
to either the detailed geometry of the circumstellar environment  or the physical mechanism by which it could form. With 
bolometric luminosities ranging from $\sim 10^4 - 10^6 L_{\odot}$   implying intial masses ranging  from  
$10M_{\odot}$ to $\geq 60M_{\odot}$  (Zickgraf et al. 
\cite{zickgraf86}, Gummersbach et al. \cite{gummersbach}, Maeder \& Meynet \cite{MM}), 
there is no guarantee that the B[e] 
phenomenon is attributable to a single physical mechanism, since stars at the extremes of 
this distribution will follow very different post-main sequence evolutionary paths. Indeed, the inclusion 
of sgB[e] stars in such schemes is  highly uncertain at present (Aret et al. \cite{aret}).

Despite their co-location in the Hertzsprung-Russell  (HR) diagram (cf. Clark et al. \cite{clark05}, 
Aret et al. \cite{aret}), it is currently unknown  whether sgB[e] stars and luminous 
blue variables (LBVs) may be physically identified with one another.
As their name suggests, LBVs are  massive evolved 
stars that are subject to pulsational instabilities that drive  dramatic photometric 
and spectroscopic variability (e.g. Humphreys \& Davidson \cite{HD}; Sect. 4.1) leading to changes in 
stellar radius,  temperature and in some cases, mass loss rate and bolometric luminosity (e.g. Groh et al. 
\cite{groh09}, Clark et al. \cite{clark09} and refs. therein). Conversely,  sgB[e] 
stars have been  considered to demonstrate little photometric variability (e.g. Zickgraf et al. 
\cite{zickgraf86}), with only a handful of known counter-examples (Sect. 4.1). Similarly, spectroscopic 
variability consistent with the canonical LBV `outbursts'  amongst sgB[e] stars has yet to be observed, 
although both classes of 
objects may show considerable overlap in spectral morphologies (e.g. Morris et al. \cite{morris}).

In order to address these issues we present new observations of the  sgB[e] LHA 115-S 18 (henceforth S18) which 
has been advanced as an LBV by a number of authors (e.g. Morris et al. \cite{morris}, van Genderen \& Sterken 
\cite{vanG02}).

\begin{figure}
\includegraphics[width=8cm,height=6cm,angle=0]{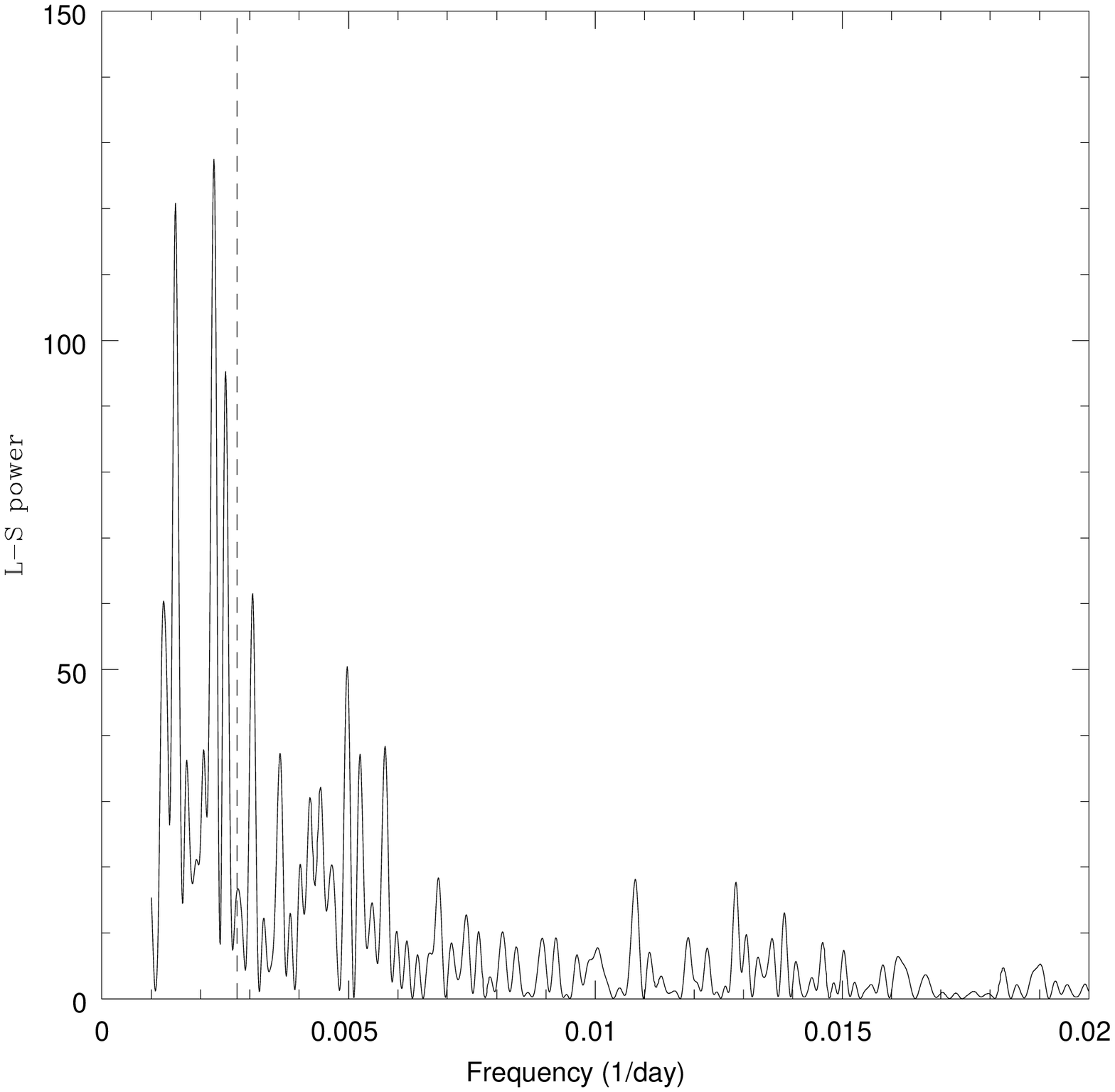}
\includegraphics[width=8cm,height=6cm,angle=0]{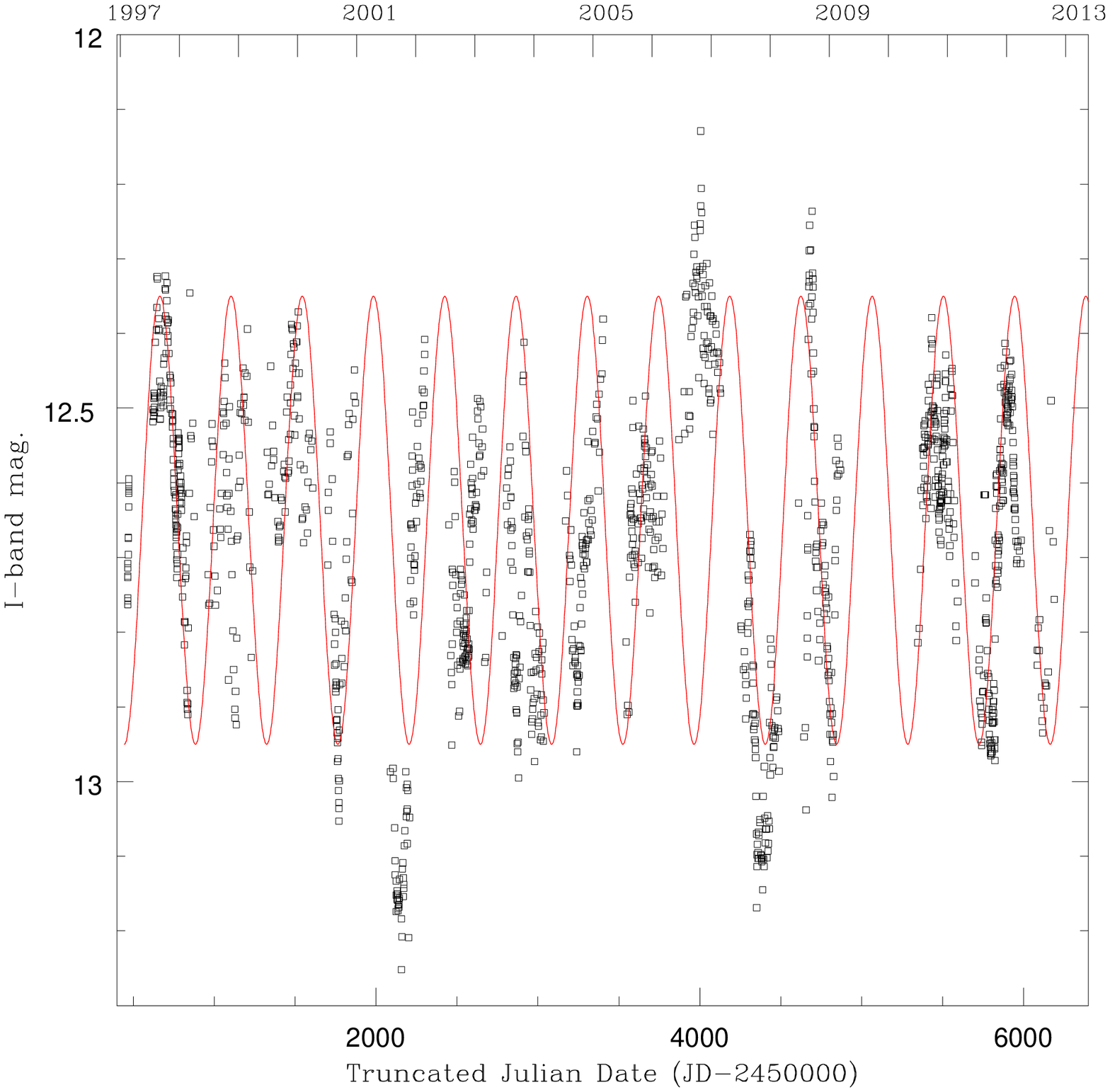}
\caption{Upper panel:  un-detrended Lomb-Scargle periodogram for the I-band photometric data, with the position of 1/year indicated by 
the dashed line. No periodic signal is returned. Lower panel: lightcurve with a sine curve corresponding to the $\sim$440~d period 
implied by the largest peak in the periodogram overplotted - the periodicity is not stable  throughout the 16~year 
duration  of observations.}  
\end{figure}

Since its discovery  by  Henize (\cite{henize}), numerous  observations have been made of S18. The optical spectrum of  the
star is characterised by strong  emission  in the Balmer series and prominent He\,{\sc i} lines (e.g. 6678{\AA} and 7065{\AA}) as 
well as 
a wealth of 
forbidden and permitted low excitation metallic lines (Fe\,{\sc ii}, [Fe\,{\sc ii}], Ti\,{\sc ii}, Cr\,{\sc ii} and O\,{\sc i}). Following 
the behaviour observed in other sgB[e] stars,
the high excitation lines are broader than the metallic lines. Moreover, they exhibit pronounced variability in profile and strength -
the Balmer lines evolve from (asymmetric) single peaked to P Cygni profiles with absorption 
extending to 750kms$^{-1}$ and vary in strength by a factor of $\sim 3$ (Zickgraf et al. \cite{zickgraf89}, Nota et al. \cite{nota}; see Table A.1 for a 
historical summary of optical 
spectroscopic 
studies).  Recently,  high resolution observations  by Torres et al. 
(\cite{torres}) revealed that the metallic lines are also variable, while  Nota et al. (\cite{nota}) 
reported changes in the radial velocities of these 
lines.  However the most dramatic variablity is seen in the He\,{\sc ii} 4686{\AA}
line, which has been observed to transition from absence to being comparable  in strength to   H$\beta$ (Table A.1), with such changes occuring 
over 
short timescales (of the order of months; Shore et al. \cite{shore87}). 

Contrary  to  the conclusions of Shore et al. (\cite{shore87}) and Zickgraf et al. (\cite{zickgraf89}),  van 
Genderen \& Sterken (\cite{vanG02}) found S18 to be photometrically variable at optical wavelengths between 1987-1991.
 Modulation occured on 
timescales from days to years, with  peak-to-peak amplitudes ranging from $\sim 0.1-1.0$~mag  (see Sect. 2.1.2. for a full description).

At other wavelengths, five International Ultraviolet Explorer  UV observations between 1981 July to 1983 March
were reported in Shore et al. (\cite{shore87}). No photometric variability above ${\sim}0.25$mag. was 
found but a number of high excitation wind features (He\,{\sc ii}, C\,{\sc iv}, N\,{\sc 
v}) were significantly variable. Two epochs of I band spectroscopy made by Shore et al. (\cite{shore87}; obtained on 1983 October 10) 
 and Aret et al. (\cite{aret}; obtained on 2005 December 10-11) share the same bulk 
morphology; being dominated by the Paschen series, Ca\,{\sc ii} 8491, 8542 and 8662{\AA} and O\,{\sc i} 
8446{\AA}. Conversely, K band spectroscopy presented by McGregor et al. 
(\cite{mcgregor}; obtained between 1987-89) and Morris et al. (\cite{morris}; obtained in 1995 November)  reveal 
evolution over this period, with the development of He\,{\sc i} 2.058$\mu$m and 2.112$\mu$m and CO bandhead emission between the two epochs. At longer wavelengths, mid-IR spectroscopy and photometry  provided by Kastner et al.
 (\cite{kastner10}) and Bonanos et al. (\cite{bonanos}) show that S18 supports a substantial IR-excess.

\begin{figure*}
\includegraphics[width=12cm,angle=-90]{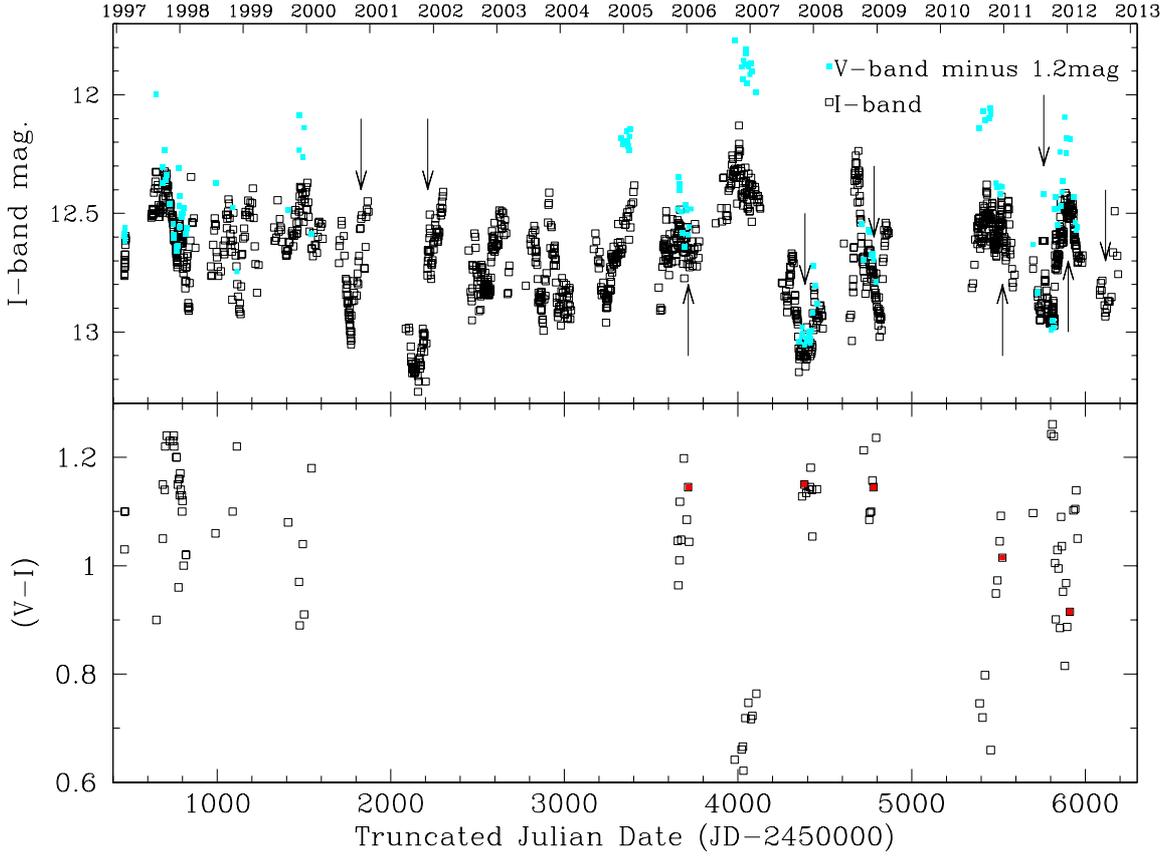}
\caption{Plots of V- and I-band OGLE lightcurves (with the V-band displaced by 1.2~mag. to enable easy comparison) and 
(V-I) colour 
index of time. Timings of optical spectroscopic observations indicated by arrows and (V-I) data within 9~d of an observation indicated by the 
red symbols (with the first four data points within 3~d). 
}
\end{figure*}

Finally a single epoch of Chandra X-ray observations  made between 2002 May-October (source $6{\_}20$ of Antoniou et al. 
\cite{antoniou}) provides a weak detection of S18. While the count rate was too low to model the spectrum, 
they estimate a flux of $\sim 3\times 10^{33}$erg s$^{-1}$ from this observation.
 Bartlett et al. (in prep.) present the results of an X-ray survey of sgB[e] stars within the Magellanic Clouds and provide
a confirmatory XMM-Newton detection from an observation on 2003 December 
18. Further details on data reduction and analysis are contained therein,
 but utilising the count rate of the EPIC-pn detector and Webpimms\footnote{Assuming $n_{\rm H}=6\times10^{20}$ cm$^{-2}$ (corresponding to the column density in the 
direction
of the SMC; Dickey \& Lockman \cite{dickey}), $\Gamma=1.7$ (similar to that adopted by the 2XMM survey; Watson et al.\cite{watson}) and a distance 
of $\sim 61\pm3$kpc (Hilditch et al. \cite{hilditch}).} it was possible to obtain a preliminary luminosity estimate of $\sim 3.9\pm0.7 \times
10^{33}$erg s$^{-1}$.

The remainder of the paper is structured as follows: we present
 our new photometric and spectroscopic observations, data 
reduction and analysis in Sects. 2 and 3.  In Sects. 4 and 5 we discuss the nature of S18 in light of these results and 
possible physical associations between sgB[e] stars, massive binaries and LBVs. Finally we summarise our 
findings in 
Sect. 6.

\section{Data Reduction}

\subsection{Photometry}

We utilized the Optical Gravitational Lensing Experiment data (OGLE-II,
-III and -IV, Udalski et al. \cite{udalski97}, Udalski \cite{udalski03}) 
to investigate the
long-term behaviour of S18 (OGLE-II designation: SMC$\textunderscore$SC6 311169). OGLE
has been regularly monitoring this object since 1997 with the 1.3-m
Warsaw telescope at Las Campanas Observatory, Chile, equipped with
three generations of  CCD camera: a single 2k$\times$ 2k chip operated in driftscan mode
(OGLE-II), an eight chip 64 Mpixel mosaic (OGLE-III) and finally a 32-chip 256 Mpixel
mosaic (OGLE-IV). Observations were collected in the standard V- and
I-bands with the vast majority of images taken through the I-band
filter.

OGLE I-band observations of S18 required an individually  tailored approach to data 
reduction. During the period of OGLE-II observations the stellar 
 brightness was well within the
photometic range of the OGLE-II camera and measurements were derived
with the standard OGLE DIA photometric reduction, based on the image difference
technique (Szyma{\'n}ski \cite{szym}). However, during the OGLE-III and OGLE-IV phases
the brightness of S18 on the survey frames turned out to be close to the
saturation level of the CCDs. In particular this happened during
observations with very good seeing conditions and as a result a few
central pixels of the star  saturated. 

Therefore in the case of OGLE-III observations we decided to employ point spread function (PSF)
photometry obtained with the DoPhot photometry program (Schechter et al.
\cite{schechter}; also provided by the OGLE-III photometry pipeline). The
PSF approach is well suited to  the treatment of  saturated
pixels at the cost of somewhat larger photometric errors.  In the case
of OGLE-IV images we manually derived S18 photometry, limiting
ourselves to the images taken during  seeing conditions where the object
was not saturated. The photometric technique was identical to the OGLE
standard photometric pipeline (Udalski \cite{udalski03}).

V-band images collected at all OGLE phases were reduced with the
standard OGLE DIA technique. S18 was far from the saturation level in
all OGLE V-band frames.

All OGLE datasets were then combined after applying second order
corrections for slightly different filter passbands. The complete
dataset comprises 1365 I-band, 133 V-band and 100 (V-I) colour index
measurements over the 16 year period from 1997-2012. Given the rapid
variability present in the lightcurve we required both V- and I-band
measurements on the same night to derive the (V-I) colour index. Errors
on the OGLE DIA photometry (I-band: OGLE-II and OGLE-IV, V-band) are
well  below 10~mmag, while those for PSF photometry (I-band: OGLE-III)
below 20~mmag. Given the much better sampled I-band lightcurve we base
our discussion on this, utilising the V-band data to provide additional
colour information.  

\begin{figure}
\includegraphics[width=6cm,angle=-90]{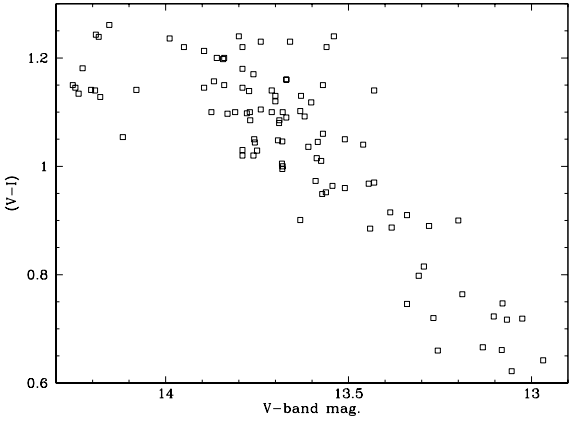}
\caption{Colour magnitude plot for S18.}
\end{figure}

\subsection{Spectroscopy}

The spectroscopic data presented in this work come from a number of sources. Firstly, we reproduce the full spectrum from
Graus et al. (\cite{graus}). This was obtained on 2010 November 24 with the MIKE echelle spectrograph on the Magellan Clay
telescope and has a resolution of $\sim 28,000$ over the 3250{\AA}-5050{\AA} wavelength range, with details on 
reduction provided in 
Graus et al. (\cite{graus}). 

Observations taken on  2011 July 20th and 21st
and 2012 July 7th were made with the fibre-fed dual-beam AAOmega spectrograph
on the 3.9~m Anglo-Australian Telescope (AAT) at the Australian
Astronomical Observatory. The instrument was operated with the Two
Degree Field ("2dF") multi-object system as front-end. Light is
collected through an optical fibre with a projected diameter of
$2\farcs1$ on the sky and fed into the two arms via a dichroic
beam-splitter with crossover at 5700\AA. Each arm of the AAOmega
system is equipped with a 2k$\times$4k E2V CCD detector and an AAO2
CCD controller. The blue arm CCD is thinned for improved blue
response. The red arm CCD is a low-fringing type. On both occasions,
the red arm was equipped with the 1700D grating, blazed at 10000\AA.
This grating provides a resolving power $R=10000$ over slightly
more than 400\AA. In the 2011 run, the central wavelength was set at
8600\AA. In 2012, it was centred on 8690\AA. The exact wavelength
range observed depends on the position of the target in the 2dF
field.

During the 2011 run, the blue arm was equipped with the 1500V
grating, which provides $R=3700$ over $\sim750$\AA. The central
wavelength chosen was 4400\AA. During the 2012 run, we used grating
580V, giving $R=1300$ over $\sim2100$\AA. The central wavelength
was set at 4800\AA.

Data reduction was performed using the standard automatic reduction
pipeline {\tt 2dfdr} as provided by the AAT at the time. Wavelength
calibration was achieved with the observation of arc lamp spectra
immediately before each target field. The lamps provide a complex
spectrum of He+CuAr+FeAr+ThAr+CuNe. The arc line lists were revised
and only those lines actually detected were given as input for  {\tt
2dfdr}. This resulted in very good wavelength solutions, with rms
always $<0.1$ pixels.

Sky subtraction was carried out by means of a mean sky spectrum,
obtained by averaging the spectra of 30 fibres located at known blank
locations. With this procedure, any existing nebular emission is not
removed.

The optical data from 2011 Decemeber 8 were taken with the ESO Faint Object Spectrograph (EFOSC2)
mounted at the Nasmyth B focus of the 3.6m New Technology Telescope (NTT), La
Silla. The EFOSC2 detector (CCD\#40)
is a  Loral/Lesser, Thinned, AR coated, UV flooded, MPP chip with
2048$\times$2048 pixels corresponding to 4.1\arcmin$\times$4.1\arcmin on the
sky. The instrument was in longslit mode with a slit width of 1.5\arcsec
using Grism 14 - resulting in a wavelength range of
$\lambda\lambda3095-5085$~\AA{} - and a grating of 600~lines~mm$^{-1}$ and a
dispersion of 1~\AA{}~pixel$^{-1}$ yielding a wavelength range of
$\lambda\lambda6000-7150$~\AA{}. The resulting spectra have a 
spectral resolution of $\sim10$~\AA{}. The data were reduced using the standard
packages available in the Image Reduction and Analysis Facility (\textsf{IRAF}).
Wavelength calibration was implemented using comparison spectra of Helium and
Argon lamps taken through out the observing run with the same instrument
configuration. 

\begin{figure*}
\includegraphics[width=15cm,angle=270]{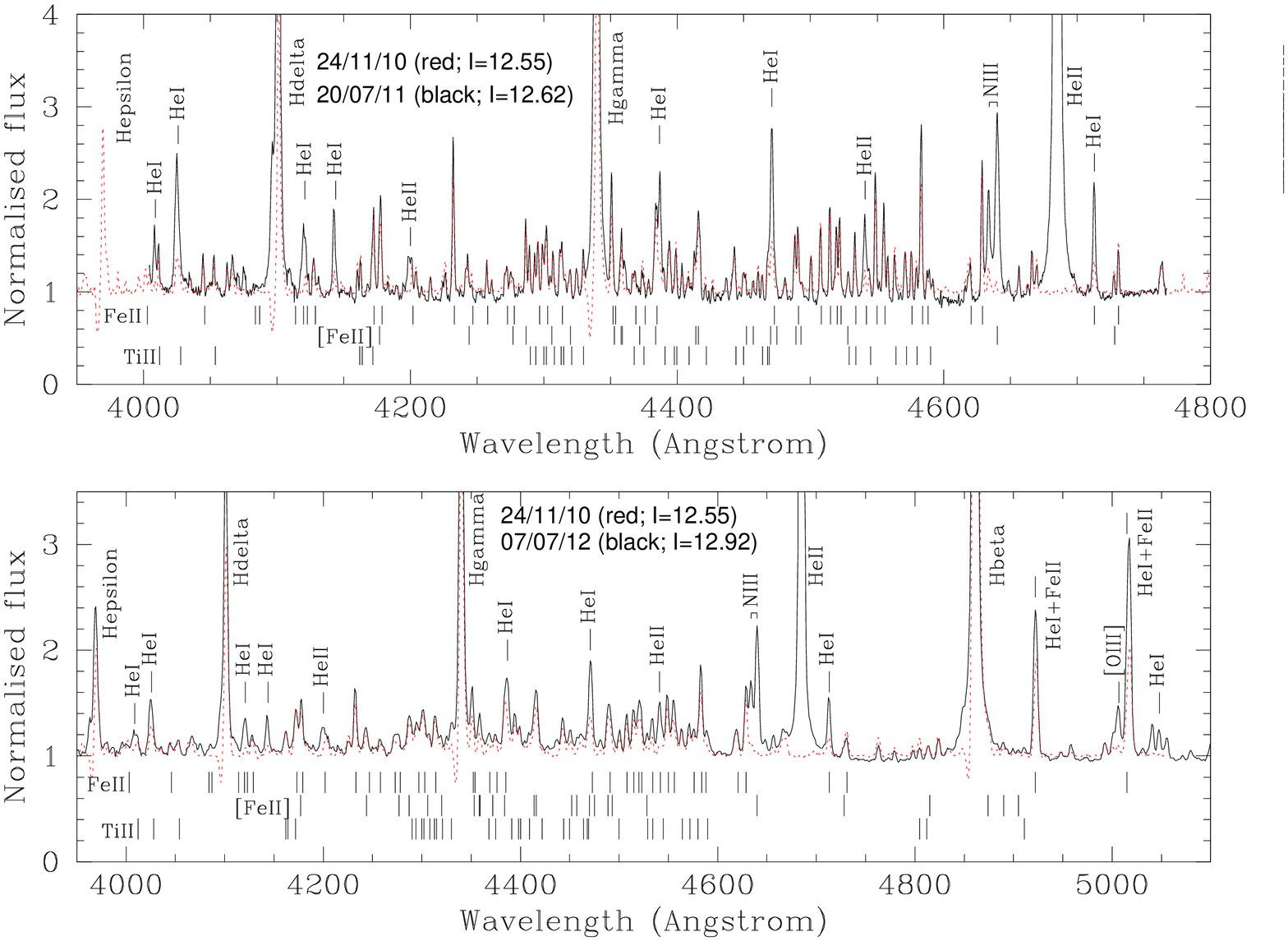}
\caption{Comparison of blue-end optical spectra obtained in 2010 November, 2011 July and 2012 July, 
after the resolution of the former has been degraded to match that of the latter two spectra (which themselves differ in 
resolution and wavelength coverage; Sect. 2).
Prominent transitions are indicated as are the I-band magnitudes on the 
dates of the observations. Following the line identifications of  Zickgraf et 
al. (\cite{zickgraf89}), additional weak emission lines from low excitation 
metallic species such as Ca\,{\sc i}, Cr\,{\sc ii}, Fe\,{\sc i}, [S\,{\sc ii}], Si\,{\sc ii} and  V\,{\sc ii} are 
not shown for reasons of clarity.}
\end{figure*}

\section{Data presentation and interpretation}

\subsection{Photometry}

Photometrically, S18 is clearly highly variable on multiple timescales (Fig. 1), mirroring the findings of van Genderen \& Sterken (\cite{vanG02}) 
and raising the possibility  
that such behaviour has continued uninterrupted throughout the $\sim22$~year period following their 1987-1991 
observations. Given this, we followed the technique outlined in 
Bird et al. (\cite{bird}) to search for periodic modulations in the I-band lightcurve. Briefly, we used the fast implementation of the
 Lomb-Scargle  periodogram  (Press \& Rybicki \cite{press}) to search for periodicities between 2~d (the Nyquist frequency for daily sampled lightcurves) 
and 10$^3$~d (above which the finite length of the lightcurve compromised sensitivity). Given the difficulties in detecting high-frequency  signals 
in the presence of low-frequency modulations we also attempted to de-trend the data. Moving in one day steps, a 
 rolling-mean was calculated over a $\sim50-300$~d time window to determine a detrending model which was then subtracted from the light curve.

No  periodicity was identified (Fig 2). Despite the appearance of various subsets  of the lightcurve (e.g. JDs$\sim 2455200-6100$) no quasi-period present 
is stable over the 16~year  duration of the observations and hence we are forced to procede via a more qualitative analysis. van Genderen
 \& Sterken (\cite{vanG02}) identified 3 types of variability: (i) short-term variability on the timescale of days and $\sim 0.1-0.2$~mag amplitude, 
(ii) $\sim 150$~d modulation with amplitude $\sim 0.25-1.0$~mag. and (iii) a long term ($\sim 2$~year) trend with amplitude $\sim 0.7$~mag upon which 
the other behaviour is superimposed. We find no evidence for the latter behaviour in our data, seeing no 
long-term trends in brightness, but can recognise the other variability described. 

Specifically, we find S18 to vary between I$\sim 12.1-13.3$~mag over the course of our observations but  cannot identify any prolonged
`quiescent' maxima or minima in this period. Rapid, low amplitude ($\sim0.2$~mag) modulation  on timescales of  
days-weeks is present - e.g. 
the periods between  JD$\sim$2453500-800, 2453900-4150 and  2455250-5600 (Fig. 1). This may be superimposed 
on the larger amplitude 
fluctuations that correspond to the second  variability type of van Genderen \& Sterken (\cite{vanG02}) - e.g. JD$\sim$2452100-300 and  2454700-800. 
The latter events span a range of timescales and amplitudes and appear exemplified by the behaviour of S18 between JD$\sim$2454000-4800 
(Fig. 1).  The  initial fade of $\sim 1$~mag amplitude - upon which is superimposed a short-lived $\sim 0.2$~mag flare -  occurs over a period of 400~d
and is followed by a $\sim 0.85$~mag brightening event of only  $\sim 50$~d duration. Note that we {\em do not} claim that the two types of photometric
variability  are the result of physically distinct processes within the system and merely employ them as 
a qualitative description of the lightcurve; indeed it is entirely possible they may both be manifestations of  
the same underlying instability. 

Interpretation of colour variability is likewise difficult and is complicated by the comparatively sparse and uneven sampling of the V-band data. 
van Genderen \& Sterken (\cite{vanG02}) report that at the photometric maxima corresponding  to $\sim$150~d variations the system appears bluer 
and we find such a broad description to be applicable to our data; the magnitude of variability appears larger in the V- rather than the I-band
 (Fig. 3).    However this behaviour appears not to be simply proportional to the brightness of the system as is clear from the colour 
magnitude plot (Fig. 4); such  a relationship is only apparent for V$\lesssim 13.6-13.8$~mag and  below this  variability  occurs at 
constant colour.  Indeed, the `blue-ing' of the system appears to be associated with episodes of brightening, with the subsequent 
decay branch of the lightcurve occuring at redder colours ((V-I)$\sim$1.2) - e.g. the `flaring' episodes between JD$\sim$2450600-800 and 2455700-6000.  
Unfortunately, better sampled data will be required to fully explore these relationships.

\subsection{Spectroscopy}

Globally, the four new blue-end spectra obtained between 2010 November and 2012 July  fit seemlessly into the long-term 
pattern observed for S18 (Table A.1 and refs. therein), although the high resolution and signal to noise (S/N) ratio of three of the 
spectra enable us to identify qualitatively new behaviour.  All four epochs are dominated by Balmer series  
emission lines, which demonstrated P Cygni profiles in all transitions in  2010 November and 
H$\gamma$ and higher lines in 2011 December  (Figs. 
5-7)\footnote{Velocities of ${\sim}500{\pm}10$kms$^{-1}$ for the blue edge of the absorption 
component in H$\beta$ and H$\gamma$ were found for the  2010 November spectrum, but were not measurable for the 
2011 December spectrum due to its lower  S/N and resolution.}.
Strong He\,{\sc ii} 4686{\AA} emission was present in two  spectra (2011 and 2012 July), 
weakly in emission in a third (2011 December) and completely absent in our highest S/N and resolution spectrum 
(2010 November). We highlight that the presence of strong He\,{\sc ii} 4686{\AA} emission appears anti-correlated 
with P-Cygni profiles in  the Balmer series in these data, with the 2011 December spectrum apparently representing a 
transitional case. Similar behaviour is visible in the  spectra from  2001 and 2005-2008 (Torres 
et al. \cite{torres}), and also appears present in earlier  observations (Table A.1).

The appearance of He\,{\sc ii} also heralds an increase in the strength of the higher Balmer series lines (Fig. 6)
and we are also able to identify a corresponding increase in strength of weaker He\,{\sc i} and He\,{\sc ii} lines 
which had hitherto been lost in the forest of low excitation metallic lines. Of particular interest are the
appearance of N\,{\sc iii} 4634{\AA} and 4640{\AA} emission lines that also accompany  He\,{\sc ii} 
4686{\AA} (Fig. 5). Conversely, while the  wealth of low excitation metallic lines\footnote{Ca\,{\sc i}, Cr\,{\sc 
ii}, Fe\,{\sc i}, Fe\,{\sc ii}, [Fe\,{\sc ii}], [S\,{\sc ii}], Si\,{\sc ii}, Ti\,{\sc ii} and V\,{\sc ii}.} 
identified by previous studies are present in all four spectra, we find little indication of variability in their
strength (cf. comparison of the 2010 November and 2011 July spectra; Fig. 5).
Likewise,  they demonstrate no appreciable evolution in width (FWHM$\sim  45\pm5$kms$^{-1}$; measured from 
the 2010 November spectrum) since the observations of Zickgraf et al. 
(\cite{zickgraf89}) and  imply much slower outflow velocities than the Balmer and He\,{\sc i} P Cygni profiles 
do. Unfortunately, comparison of the line widths of the low excitation species
 between  epochs is  not 
instructive due to the lower resolution of our remaining spectra, but we see no appreciable differences in radial 
velocity in our data.

We therefore arrive at a simple picture where the behaviour of high- and low excitation lines appears largely 
de-coupled, with the former showing dramatic variations in line profile and strength over timescales as short as 
months, while the latter remain largely unaffected. Moreover, this pattern appears qualitatively consistent with 
observations made over the past half century (Sect. 1.1 and Table A.1). 

The  observations available in the  6000-7200{\AA} and 8400-8900{\AA}  windows likewise  
indicate that the overal spectral morphologies remain largely unchanged from previous epochs (Sect. 1.1; Figs. 7 
\& 8).  Specifically, and as expected, H$\alpha$ and the Paschen series and a number of He\,{\sc i} lines  are strongly in emission, although 
the I-band spectrum is dominated by emission from Ca\,{\sc ii} and O\,{\sc i}. Other low excitation species present 
and in emission are Fe\,{\sc ii}, Si\,{\sc ii} and N\,{\sc i}. Forbidden [O\,{\sc i}] 6300{\AA} line is also in emission; 
while consistent with a [N\,{\sc ii}] identification we 
suspect the inflection in the blue flank of H$\alpha$ is instead due to the run of absorption in the stellar wind 
(also seen in other sgB[e] stars such as R126; Zickgraf et al. \cite{zickgraf86}). Finally we identify TiO bandhead 
emission, as has previously been observed (Zickgraf et al. \cite{zickgraf89}).  

Unfortunately, the low excitation emission lines (e.g. [O\,{\sc i}], [O\,{\sc ii}], Ca\,{\sc ii} and [Ca\,{\sc 
ii}]) employed by Kraus et al. (\cite{kraus10}) and Aret et al. (\cite{aret}) to determine disc kinematics are 
either absent, outside our coverage, or 
observed at too low a resolution to impart information on the velocity structure of the emitting 
region. However, in the absence of the  shell-type emission line profiles that arise from  obscuration of the 
stellar disc by an 
intervening circumstellar disc (cf. S65; Kraus et al. \cite{kraus10}) we infer the system is not observed 
edge-on.

\subsection{Photometric/spectroscopic comparison}

We may utilise our well sampled optical lightcurve to place the spectroscopic observations into context 
(Figs. 1 \& 3); noting that unfortunately
the latter sample a  restricted photometric  range (I$\sim13.10 - 12.51$~mag), albeit including  local maxima and 
minima\footnote{van Genderen \& Sterken (\cite{vanG02}) report that the spectrum presented by Zickgraf et al. (\cite{zickgraf89}) - in which He\,{\sc ii} 
4686{\AA} was absent - was obtained during a photometric maximum with V$\sim$13.3; comparable in magnitude to the peak in our lightcurve 
around JD$\sim2455000$.}.
 Nevertheless, the data {\em suggest} that the changes in spectral morphology - and in particular the appearance of He\,{\sc ii} 4686{\AA}  emission - 
is {\em not} uniquely correlated with photometric magnitude (Table A.1). 
This apparent lack of 
 correlation is exemplified by the spectra obtained in 2010 November  and 2011 July, where the strength of the 
 high excitation emission features in the spectra 
differ greatly  (Figs. 5 and 6) but for which the I-band magnitudes are identical to within $\sim$0.07~mag. At the times of the two 
observations   the extended photometric behaviour of the star was rather different, respectively 
exhibiting rapid low amplitude variability and at the peak of a 
rapid ($\sim 60$~d) $\sim$0.35mag `flare' (Fig. 1), which  might be supposed to be a cause of the different 
spectral morphologies. 
However, He\,{\sc ii} 4686{\AA} emission is present in both the 2011 and 2012 July spectra - obtained in local photometric minima  and maxima - implying
 a lack of a clear correspondence between lightcurve behaviour and spectral morphologies. 
Nevertheless,  our limited dataset is consistent with the  possibility that while 
He\,{\sc ii} 4686{\AA} emission may occur in any photometric state, a lack of emission is only associated with brighter states.
Unfortunately, with only 5 quasi-contemporaneous observations (Fig. 3) we are unable to draw any conclusions regarding correlations to 
the (V-I) colour; again,  more frequent spectroscopic sampling will be required to explore any such 
behaviour.

\begin{figure}
\includegraphics[width=9cm,angle=0]{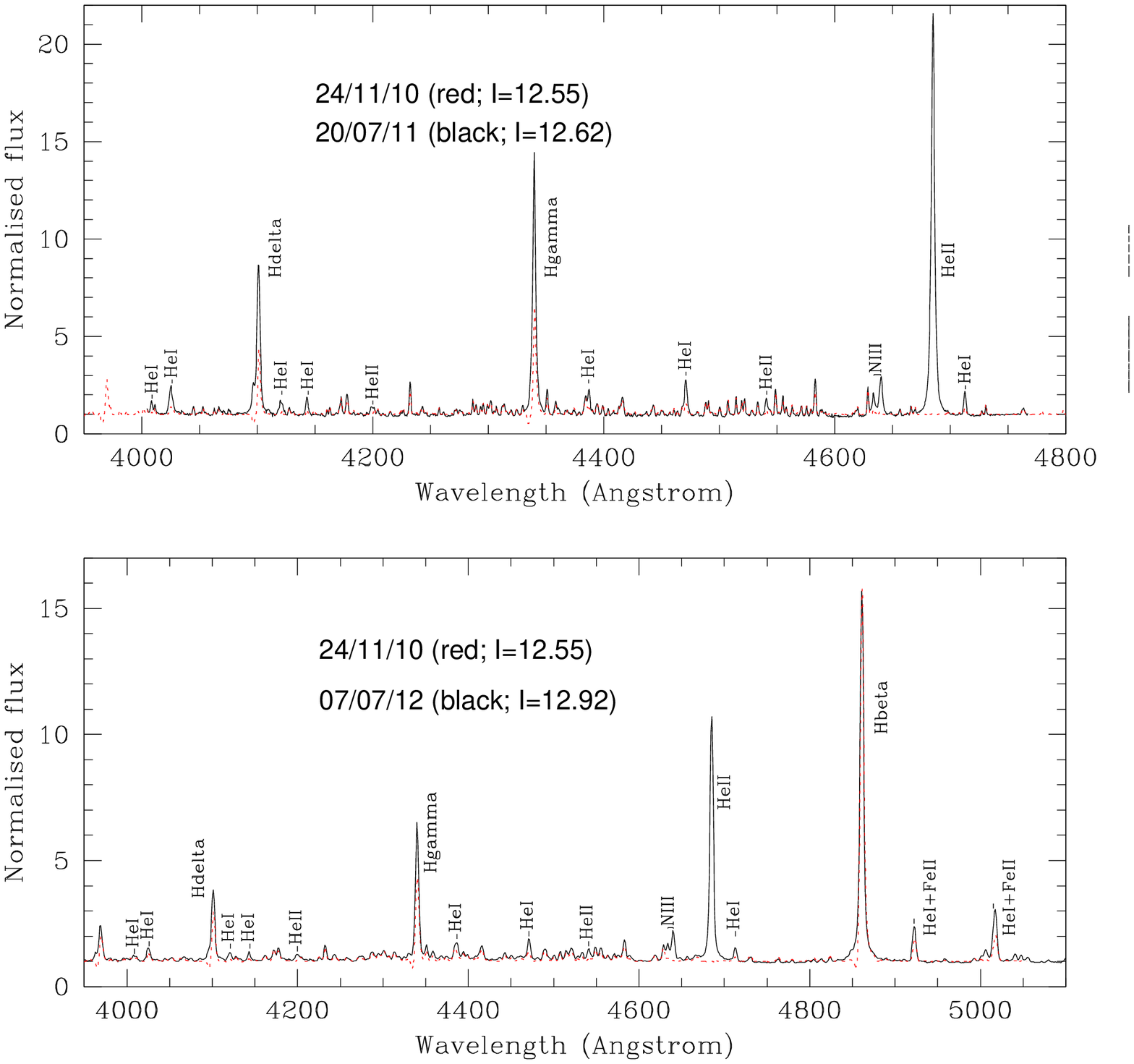}
\caption{As Fig. 5 but spectra scaled to compare the stronger emission features.}
\end{figure}

\begin{figure}
\includegraphics[width=9cm,angle=0]{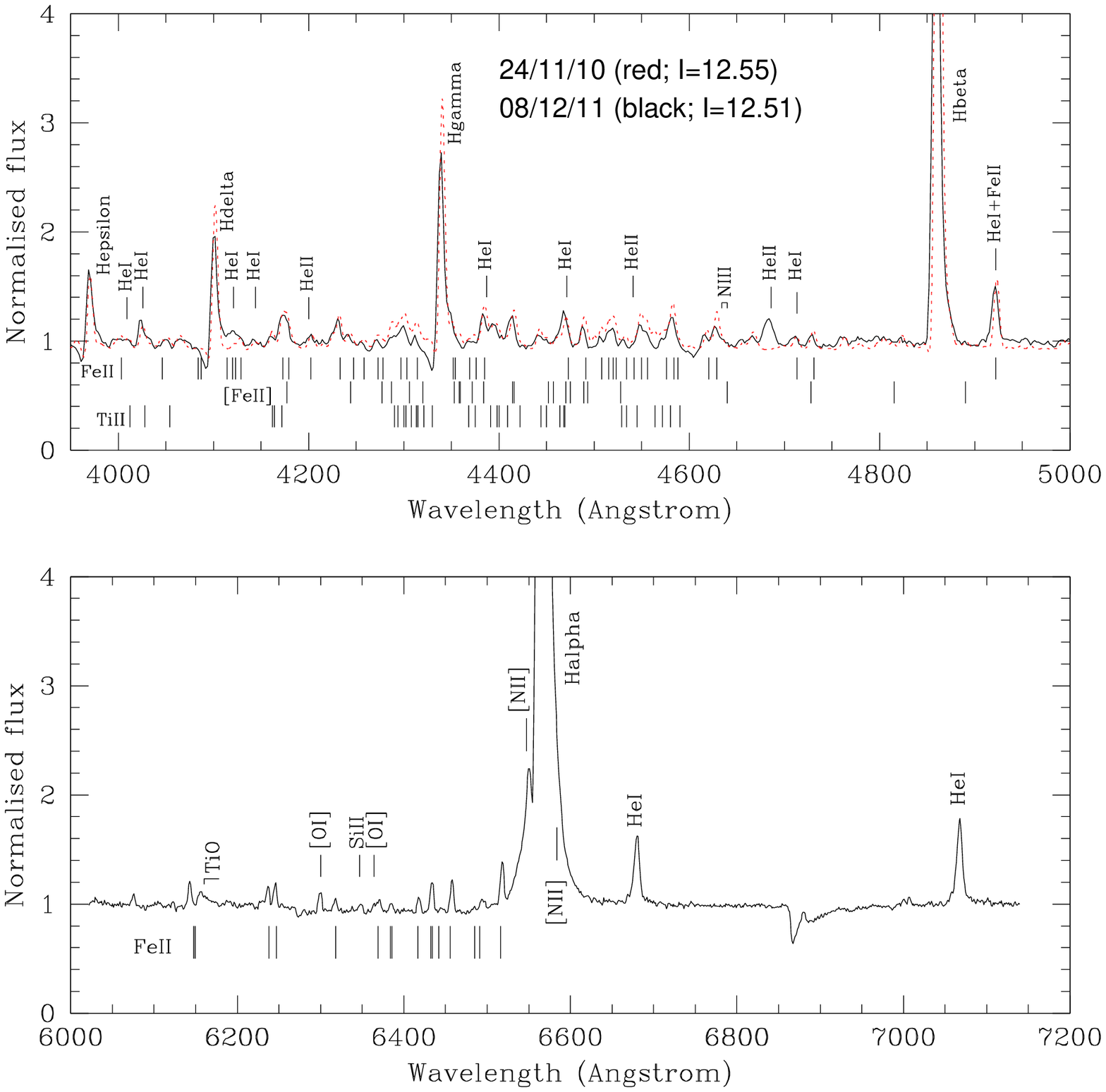}
\caption{As Fig. 5 but with the low resolution spectrum from 2011 December 8 - the only epoch for 
which coverage 
extends 
to H$\alpha$ and   the He\,{\sc i}
6678 and 7065{\AA} transitions. Note the peak intensity of H$\alpha$ is $\sim 35 \times$ continuum at this time.}
\end{figure}

\begin{figure}
\includegraphics[width=7cm,angle=270]{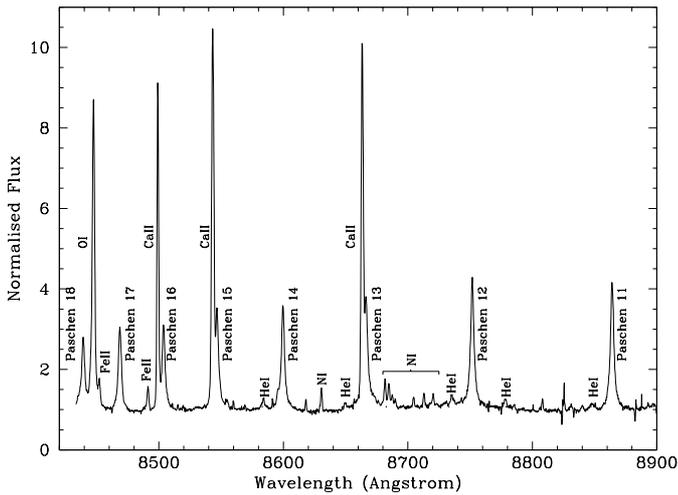}
\caption{I-band spectrum of S18 from 2011 July 20 with major transitions indicated.}
\end{figure}

\section{The circumstellar environment of S18}

So how is S18 to be understood in the context of other sgB[e] stars? A defining property of such stars is the presence of
a cool equatorial disc or  torus, comprising  neutral and molecular gas and warm dust.  The presence of e.g. Ca\,{\sc ii} emission (Fig. 8)
implies cool, dense conditions ($\sim 5000$K, $\sim 10^9$~cm$^{-3}$; Hamann \& Simon \cite{hamann}) which are shielded from the ionising flux of the 
central star. Likwise O\,{\sc i} 8446{\AA} requires a dense transitional zone between H\,{\sc i} and H\,{\sc ii} regions 
($10^9 - 10^{10}$cm$^{-3}$; Grandi \cite{grandi}) for Ly$\beta$ pumping to drive it strongly into emission (cf. Wd1-9; Clark at al. 
\cite{clark13}) and the discovery of Raman-scattering lines also requires neutral hydrogen to be present  (Torres et al. 
\cite{torres}).  The presence of TiO and  CO bandhead emission (Sect. 1.1 and Fig. 7) likewise points to the presence of cool,
dense molecular gas (Morris et al. \cite{morris}, Liermann et al. \cite{liermann}). At longer wavelengths the 
mid-IR properties of S18 suggest the presence of warm 
($\sim$800K) dust with a probable crystaline silicate composition; both 
properties consistent with other sgB[e] stars (Kastner et al. \cite{kastner06},
\cite{kastner10}, Bonanos et al. \cite{bonanos}).

In these regards S18 appears to conform to the stereotype of a sgB[e] star (Zickgraf et al. \cite{zickgraf85}), where the low excitation spectrum is 
formed within - or in the case  of  the Ly$\alpha$ and $\beta$ pumped lines such as O\,{\sc i} 8446{\AA} and Fe\,{\sc ii} 8490{\AA} - on the 
surface layer of 
a cool dense torus  surrounding the  central star. The optical-IR continuum excess demonstrated by S18 (Zickgraf et al. \cite{zickgraf89}) would
 then result from a combination of  thermal  bremsstrahlung from the stellar wind and/or disc and thermal emission from the warm dust which 
also resides in the torus (e.g. Porter \cite{porter}).  

Recent spectroscopic observations suggest that the neutral gaseous component in sgB[e] stars is detached from the central 
star  (e.g. Kraus et al. \cite{kraus10}, Liermann et al. \cite{liermann} \&
Aret et al. \cite{aret}). Kastner et al. 
(\cite{kastner06}, \cite{kastner10}) arrived at a similar  conclusion from analysis of the mid-IR properties of sgB[e] stars and, by analogy to lower mass AGB stars, postulated that the dust resides in long lived quasi-Keplerian disc surrounding  a 
binary system which  formed from mass lost in a previous common-envelope/mass transfer
phase.
Support for this geometry has been provided by interferometric observations of the galactic B[e] stars CI Cam and MWC300
 (Thureau et al. \cite{thureau}, Wang et al. \cite{wang}) and the sgA[e] star HD 62623 (Meilland et al. 
\cite{meilland}), which imply ring-like geometries for the circumstellar material.

While our spectroscopic observations lack the wavelength coverage and resolution to address the kinematics of the 
disc/torus, the spectra of S18 presented by Aret et al. (\cite{aret}) reveal that the higher excitation 
[Ca\,{\sc ii}]
lines are broader than the lower excitation [O\,{\sc i}] lines, as anticipated for Keplerian rotation.

Consequently, it is possible to see why the low-excitation component of the spectrum of S18 is, in terms of 
variability,
 essentially decoupled from the high-excitation component, {\em if} the former arises in a largely static,  
quasi-Keplerian, viscous, detached disc while 
the latter is associated with an intrinsically variable  star.
Following the analysis of classical Be star discs by Okazaki et al. (\cite{okazaki}) we may obtain an 
 estimate of the viscous timescale of such a  disc.
We make the simplifying assumption that the gaseous component of the quasi-Keplerian 
disc is isothermal and adopt a representative disc temperature of 
$T_{\rm d} \sim 6000$K\footnote{The mid-point Kraus et al. (\cite{kraus10}) derive from the strength of [O\,{\sc i}] emission lines, noting that as with classical Be stars the assumption of an iothermal disc is likely violated in 
practice.} and adopt representative  
inner disc radii of  $R_{\rm d} \sim R_{\ast}$ (i.e. it extends to the stellar surface at  $R_{\ast}$) and  
$R_{\rm d} \sim 10R_{\ast}$ (motivated by the apparent truncation of the disc in  LHA 115-S 65; Kraus et al. \cite{kraus10}, Aret et al. \cite{aret}).

Under these conditions the  viscous timescale in the disk, $t_{\rm vis}$, is given by
$(\alpha \times (H/r)^2 \times \Omega_{K(r)})^{-1}$, where $H$ is the disk scale-height and
$\Omega_{K(r)}$ is the Keplerian angular frequency at radius $r$. Adopting  $M_{\ast}\sim 19M_{\odot}$ and 
$R_{\ast}\sim 33R_{\odot}$ (e.g. Aret et al. \cite{aret}) for $r=R_{\ast}$
we have   $\Omega_{K} \sim 0.8$~d and  $H/r \sim {c_s}/(r \times \Omega_K) \sim 0.02$ and 
$\Omega_K \sim 25$~d and  $H/r \sim 0.06$ for $r=10R_{\ast}$ (where $c_s$ is the sound speed).
From these values we find $t_{\rm vis} \sim 20/({\alpha}/0.3)$~yr and $\sim 62/({\alpha}/0.3)$~yr respectively 
(where $\alpha$ is the viscosity parameter and $t_{vis}$ is proportional to the inverse of the disc 
temperature).  Therefore the combination of the relatively low temperature  (in contrast to classical Be star 
discs) and large inner radius adopted results in a long viscous timescale and hence timescale of variability 
in response to changes in the bulk properties of the disc/torus. 

Despite this, both Nota et al. (\cite{nota}) and Torres et al. (\cite{torres}) report line profile variability in the 
Fe\,{\sc ii} lines - one  might expect rapid changes in the disc properties in response to changes in the ionising flux 
or, by analogy to classical Be stars, due to the impulsive injection of matter (Rivinius et al. \cite{rivinius}).

\section{The nature of the central star(s)}

\begin{figure*}
\includegraphics[width=15cm,angle=270]{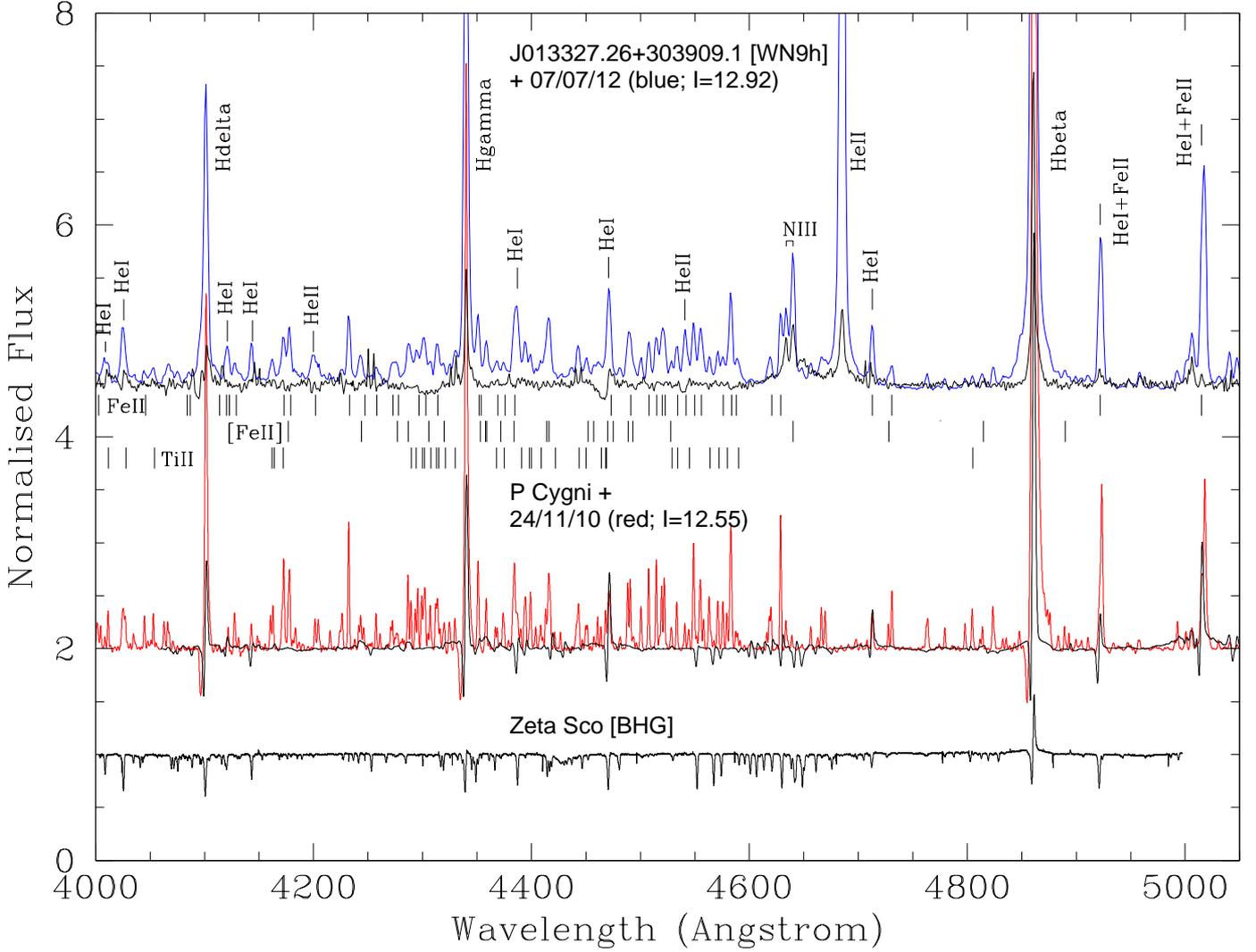}
\caption{Comparison of 2010 (`cool' state) and 2012 (`hot' state) spectra versus the B1 Ia$^+$ hypergiant $\zeta^1$ Sco, P Cygni and the M33 WN9h star 
J013327.26+303909.1 with major transitions indicated. Note that in the absence of a cool, toroidal circumstellar envelope, we
would not expect the three comparison spectra to demonstrate the rich low-excitation metallic emission spectrum of S18.}
\end{figure*}

\begin{table*}
\begin{center}
\caption[]{Stellar parameters of comparison stars in Fig. 9}
\begin{tabular}{lcccccc}
\hline
\hline
Star & log($L/L_{\odot}$) & $T_{\rm eff}$ & \.{M} & ${\varv}_{\infty}$ & BC & Ref. \\
     &                    &  (kK)     & ($10^{-6} M_{\odot}$yr$^{-1}$) & (km s$^{-1}$) & (mag) & \\
\hline
$\zeta^1$ Sco (B1 Ia$^+$) & 5.93 & 17.2 & 1.55 & 390 & -1.6 & 1 \\ 
P Cygni (B1 Ia$^+$/LBV)   & 5.85 & 19.0 & 30.0 & 185 & -1.8 & 2 \\
Mean WN9h                 & 5.7  & 27.6 & 18.5 & 450 & -2.9 & 3 \\
\hline 
\end{tabular}
\end{center}
Tabulation of the bolometric luminosity ($L$), effective temperature ($T_{\rm eff}$), mass loss rate (\.{M}), terminal
wind velocity (${\varv}_{\infty}$) and bolometric correction (BC) for the comparison spectra in Fig. 9. References are $^1$Clark et al. (\cite{clark12b}), $^2$Najarro (\cite{paco}) and $^3$Crowther \& Smith (\cite{crowther}); the value for the final 
entry is the mean of the modeling results for the four WN9h stars analysed by the latter authors.  The broadly comparable 
luminosities reveal that the changes in spectral morphology are primarily the result of changes in temperature and wind 
parameters. 
\end{table*}

What is the nature of the hot, luminous central star in S18? No uncontaminated photospheric features are present in any 
published spectra so we are forced to infer stellar properties from the highly variable high-excitation emission-line 
spectrum. Following the suggestions of Zickgraf et al. (\cite{zickgraf89}) and Nota et al. (\cite{nota}), in Fig. 9 we plot 
examplar spectra from November 2010 (henceforth the `cool' spectrum/phase) and July 2012  (the `hot' spectrum/phase)
 against representative spectra of the  early blue hypergiant (BHG) $\zeta^1$ Sco, the BHG/LBV P Cygni and a sample WN9h star,  
J013327.26+303909.1.
As described by these authors, in both cool and hot states S18 shows gross similarities  to the latter two objects. 
Specifically, (i) a simple spectrum comprising narrow Balmer and He\,{\sc i} emission lines in the cool phase, with the 
former showing P Cygni profiles and (ii) the disappearance of P Cygni line profiles and the 
development/strengthening of 
He\,{\sc i}, He\,{\sc ii} and N\,{\sc iii} emission in the hot phase; all characteristic features of the spectra of 
P Cygni-type LBVs and WN9h stars, respectively (e.g. Najarro \cite{paco}, Crowther \& 
Smith \cite{crowther}). Neither  the hot nor the cool spectrum resembles that of the early BHG $\zeta^1$ Sco and, by extension,
 field blue supergiants, because such stars do not support winds of sufficiently high mass loss rate 
 to drive the Balmer series into the  intense emission present in S18
 (Table 1 and Clark et al. \cite{clark12b} and refs. therein). 

Under the assumption that the high-excitation emission line spectrum arises in a stellar wind,  
the blue edge of the P Cygni profiles in the Balmer series provides a 
lower limit to  the  wind 
velocity in the cool phase ($\sim 500-750$kms$^{-1}$). Such a value is 
  broadly consistent with expectations for early-mid  B supergiants (BSG)/BHGs, but much faster than 
expected for LBVs such as P Cygni (Table 1 and Clark et al. \cite{clark12b}). Conversely the mass loss rate 
($\sim 3 \times 10^{-5} M_{\odot}$yr$^{-1}$; Zickgraf et al.  \cite{zickgraf89},
derived from line profile modeling of the Balmer series) is much higher than 
expected for BSGs and BHGs but  consistent with known LBVs (Clark et al. 
\cite{clark12b})\footnote{Nevertheless, this pattern of wind properties is observed in the unusual BHG Wd1-5; a star which 
is thought to be the result of binary driven mass loss and that is also thought to approach the Eddington limit (Clark et al. \cite{science}).}.
The disappearance of P Cygni profiles in  the  WN9h spectrum is explicable in terms of the 
full ionisation of the stellar wind due to an increase in stellar temperature,  although one might also expect such behaviour to be 
accompanied by changes in the stellar wind;  note, for example, the greater wind velocity  of WN9h stars with respect to P 
Cygni-type LBVs (for similar mass loss rates; Table 1).
 Under such a picture the spectrum obtained in 2011 December, which 
demonstrates weak  He\,{\sc ii} emission and P Cygni profiles in the 
higher Balmer transitions, would represent an intermediate state. 

An obvious interpretation of these spectroscopic data is that S18 transitions directly between P Cygni-type LBV and WN9h spectral types. 
However there are several {\em caveats} to such an hypothesis.
 Firstly,  Zickgraf et al. (\cite{zickgraf89}) presented a simple model atmosphere fit to the UV-IR spectral 
energy distribution of S18 and, assuming a temperature of $\sim$25000K, a bolometric correction of -2.5mag
and a distance to the SMC of 60kpc, inferred a luminosity of  log($L/L_{\odot}) \sim 5.6$ in the cool phase. The 
lower bolometric correction derived for P Cygni  suggests a downwards revision to  log($L/L_{\odot}) \sim 5.3$;
significantly lower than in the comparison stars  listed in Table 1 (although low luminosity examples are known  - e.g. Clark et al. 
\cite{clark09b}).

Secondly, in both cool and hot phases the emission line spectrum is unprecedentedly strong  with respect to 
P Cygni-type LBVs and WN9h stars; this being   particularly noteworthy  for   He\,{\sc ii} 4686{\AA} (Fig. 9: Nota et al. 
\cite{nota}).  Moreover,  He\,{\sc ii} 4686{\AA}   was  seen to be double peaked at one epoch, with the blue peak of
 greater intensity (with  corresponding   asymmetric profiles in the 
Balmer series 
and He\,{\sc i}; Nota et al. \cite{nota}). Such asymmetries are observed in other B[e] stars (e.g. Zickgraf 
\cite{zickgraf03}) and are also common in classical Be stars (Okazaki \cite{okazaki}); in all cases 
heralding a departure from spherical symmetry for the circumstellar environment, as we infer for  S18.
This is in contrast to the spherical  geometry typically assumed for the 
the winds of P Cygni-type LBVs and WN9h stars (although Nota et al. (\cite{nota}) report similar asymmetries for a
subset of the Ofpe/WN9 stars they studied).  Another unexpected 
spectral peculiarity in S18  is the absence of P Cygni profiles in the He\,{\sc i} emission lines in the cool spectrum when 
clearly present in the Balmer series. 

Finally, the X-ray detection ($L_{\rm X} \sim 3\times 10^{33}$erg s$^{-1}$; Antoniou et al. \cite{antoniou}, Bartlett et al. in prep.) 
strongly suggests a binary nature for S18. The  canonical X-ray luminosity for a single massive star 
($L_{\rm X} \sim 10^{-7} L_{\rm bol}$; Berghoefer et al. \cite{berghoefer}) implies $L_{\rm X} \sim 1.6\times 10^{32}$erg s$^{-1}$ 
 for S18; over an order of magnitude lower than observed (Sect. 1.1) and hence  requiring an additional 
source of X-rays to be present within the system. Therefore in addition to a contribution from the circumstellar disc/torus one might
also suppose additional spectroscopic contamination from a binary companion; we return to this below.

\subsection{S18 as a LBV}
With these caveats in mind,  the simplest explanation for the spectrosopic 
and photometric properties of S18 remains  that it hosts an LBV.
Indeed, a  physical link between sgB[e] stars and LBVs has been proposed on the 
 basis of photometric variability in a number of stars\footnote{The binary RMC 4  (Zickgraf et al. \cite{zickgraf96}), LHA 120-S 22 (Shore et al. \cite{shore90},\cite{shore92};
van Genderen \& Sterken \cite{vanG99}; Sterken \cite{sterken}) 
LHA 120-S 134 (Stahl  et al. \cite{stahl84}, van Genderen {vanG01}) and J013459.47+303424.7 (Clark et al. \cite{clark12a}) are all reported as photometrically variable at  UV-optical wavelengths, albeit from 
considerably sparser datasets than considered here.}.

S18 appears to reside in a region of the H-R diagram prone to pulsational instabilities (e.g. Cantiello et al. \cite{cantiello}, Saio 
\cite{saio}) and clearly shows the presence of the rapid microvariability present in many LBVs and 
  luminous super-/hypergiants (the $\alpha$-Cygni variables; van Leeuwen et al. \cite{vanL})\footnote{Following the analysis of the blue supergiant
HD~50064 by Aerts et al. (\cite{aerts}), an identification of these variations with strange mode oscillations would appear an interesting hypothesis, especially 
given the long-term changes in both amplitude and timescale of this behaviour. However, the likely progenitor mass of S18 ($\sim 20-25 M_{\odot}$ via 
comparison
to the isochrones of Groh et al. \cite{groh13}) is lower than that expected for stars encountering this instability ($\sim 40 M_{\odot}$; Aerts et al. 
\cite{aerts}).}
but the rapid, high-amplitude 
variability it demonstrates appears at variance with typical canonical LBV excursions (e.g. Spoon et al. \cite{spoon})\footnote{These consist of 
 (i) microvariations, occuring over weeks to months and of low amplitude ($\lesssim 0.2$~mag),
 (ii) LBV excursion, occuring over years to decades and typically of $\sim 0.5-2.0$~mag amplitude and which
 are typically  associated with colour changes in the sense that the star becomes redder as it brightens and 
(iii) giant eruptions, in which the star brightens by $\gtrsim 2.0$~mag over timescales ranging from 
days to years and in which the bolometric luminosity is not conserved (e.g. Humphreys \& Davidson \cite{HD}, Lamers et al. 
\cite{lamers}).}. Nevertheless such rapid variability is not completely unknown, being 
present in a handful of local LBVs and also the extragalactic SNe imposters, presaging both giant eruptions ($\eta$ Carinae, HD5980 and 
 SN 1954J; Frew \cite{eta}, Smith \& Frew \cite{frew}, Breysacher \cite{breysacher97} and Tammann \& Sandage 
\cite{tammann}) and {\em bona fide} SNe
(SN 2010mc and possibly SN2009ip; Ofek et al. \cite{ofek} and Sect. 6), although in the case of 
SN2000ch a decade of such behaviour has yet to lead
to either outcome (Pastorello et al. \cite{ch}). S18 may be identified with  SN2000ch in this regard, although the amplitude of variability 
is significantly lower, most closely resembling   SN 1954J in the 4 years leading up to its eponymous outburst (Tammann \& Sandage \cite{tammann}; see also SN2009ip 
(Sect.6)). 

The colour behaviour of S18 also differs from that of LBV excursions; leading to the system becoming bluer in the rising 
branches of the lightcurve and showing much greater variance in comparison to canonical examples (cf. Fig. B.1, derived
from data presented by Spoon et al. \cite{spoon}). The colour behaviour of normal LBVs - which become redder as they brighten - arises from the simultaneous 
expansion (contraction) and cooling (warming) of the photosphere as the star transits from WN10-11h to cool supergiant in 
`outburst' and back again (e.g. AG Car; Groh et al. \cite{groh09}, Groh priv. comm. 2012) and clearly cannot explain 
the behaviour observed here.  Indeed the spectral evolution of  S18 occurs over a temperature range for which 
 the change in (V-I) colour index would be minimal (cf. Clark et al. \cite{clark05b}). 

We highlight that not all LBVs show this behaviour - for 
instance the SNe imposter SN 2000ch shows no colour/magnitude correlations during its rapid high amplitude flares 
(Fig. B.1; Pastorello et al. \cite{ch}).
An additional complication with interpreting the behaviour of  S18 is that the presence of a substantial 
circumstellar envelope - and  a probable  binary campanion - may also contribute to the photometric variability; indeed the 
modeling of Zickgraf et al. (\cite{zickgraf89}) suggests a continuum excess extending to optical wavelengths.

We may advance two different scenarios to explain the photometic variability, noting that they are not mutually exclusive. 
Firstly, the  episodes in which S18 rapidly brightens and becomes bluer may be due to an increased contribution 
from the hot central star relative to the cool circumstellar disc/torus (and/or stellar wind) in the combined emergent 
spectrum.  The lack of colour changes associated with  `quiescent' and fading branches of the lightcurve is more difficult to 
understand, but may represent quasi-simultaneous variability in  both emission components. Moreover, we are unable 
to advance a physical mechanism for such stellar `pulsations', although this  is true for  LBVs in general; for instance the pulsations 
Aerts et al. (\cite{aerts}) attribute to strange mode oscillations in HD 50064 are of substantially lower magnitude ($\sim0.2$mag) to those 
observed here.

Alternatively the variability may be  due to changeable circumstellar  extinction, with comparable  behaviour seen in both 
classical Be  (de Wit  et al. \cite{dewit}) and Herbig Be stars (Lamers et al \cite{lamers99}). In classical Be stars this
 is observed for edge-on systems, where the gaseous disc projected against 
the star reduces continuum 
emission at short wavelengths, but  contributes continuum emission at  predominantly longer 
wavelengths from the portion of the disc {\em not} projected against the 
star, such that the system becomes fainter and redder as the disc optical 
depth increases. A variant of this  scenario is assumed for 
Herbig Be stars, where variable stellar extinction is  attributable to a clumpy, dusty circumstellar envelope. 

However, a significant problem with such an hypothesis 
is the location of the obscuring material and the rapidity of changes. A comparison of the line profiles of S18 to e.g. those of the 
sgB[e] star LHA 115-S 65 suggest it is not seen edge-on (Zickgraf et al \cite{zickgraf89}, Kraus et al. \cite{kraus10}); therefore it appears
unlikely that the circumstellar disc/torus could be responsible as is assumed for both classical and Herbig Be stars. Likwise, while the stellar winds of
LBVs are known to be clumpy (Davies et al. \cite{davies}), such clumps are not expected to  be sufficiently dense to provide  the 
requisite obscuration of the stellar disc.

Irrespective of the physical mechanism, a key observational finding of this research is that there is no clear correlation between
 photometric and spectroscopic variability;  for example the hot and cool phase spectral states in 2010 November and 2011 July 
differed by only $\sim 0.07$mag. Naively 
applying  the respective bolometric  corrections implied by the spectral morphologies  (Table 1) would imply that the luminosity of 
S18 had increased by $\gtrsim 0.44dex$ over this period;  however such a conclusion is tempered by the potentially highly 
variable continuum  contribution from the stellar wind/circumstellar envelope at these wavelengths. Bearing  this important {\em 
caveat} in mind, changes in bolometric luminoity are not unprecendented in {\em bona fide} LBVs, having been 
observed in e.g. AFGL 2298 and Romano's star (Clark et al. \cite{clark09}, Maryeva \& Abolmasov \cite{maryeva}). Moreover, while S18 has not
 been  observed to transition to the  cool hypergiant phase that characterises many LBV excursions, the range of spectroscopic variability it exhibits 
(WN9h$\Leftrightarrow$P Cygni-type LBV) mirrors the behaviour of Romano's Star (Maryeva \& Abolmasov \cite{maryeva}). 

Finally, we note that the mid-IR properties of S18 appear to exclude the presence of the cool dust that characterises the circumstellar ejecta of many
 LBVs (Clark et al. \cite{clark03}, Bonanos et al. \cite{bonanos}). 

\subsection{S18 as a binary system}
The presence of highly variable He\,{\sc ii} 4686{\AA} emission in the spectrum of S18 has long been attributed to binarity (Shore et al. \cite{shore87}, Zickgraf et 
al. \cite{zickgraf89}). Furthermore Naz\'{e} et al. (\cite{naze12}) find that in the absence of a binary companion LBVs are intrinsically X-ray faint, 
strongly suggesting that our X-ray  detection  of S18 indicates the presence of an unseen binary companion (Sects 1.1 and  4.0).  Two options therefore present themselves; accretion onto a compact object or wind 
collision leading to shocked material in a massive binary. 

Several examples of X-ray binaries with sgB[e]-like mass donors are known. SS433 has a complex, dusty  circumstellar environment (Clark 
et 
al. \cite{clark07}) and demonstrates highly variable He\,{\sc ii} 4686{\AA} emission (Fabrika \& Bychkova \cite{fabrika}),
 but is several orders of magnitude brighter at X-rays than S18
(Cherepaschuk et al. \cite{cher}).  CI Cam and IGR J16318-4848 appear to provide better comparisons; while distances to both stars are 
uncertain, at the lower end of the ranges  quoted in the literature 
 their quiescent fluxes are comparable to that of S18 (Bartlett et al. 
\cite{bartlett}, Barrag\'{a}n et al. \cite{barragan}). 

Spectroscopically, CI Cam  also bears a close 
resemblance, even to the extent of 
He\,{\sc ii} 
4686{\AA} emission being present in quiescence  (e.g. Clark et al. \cite{clark99}, Hynes 
et al. \cite{hynes}). 
Nevertheless, the long term photometric and spectroscopic behaviours of CI Cam 
 and S18 are very different, with the former much less variable and demonstrating consistently weaker 
He\,{\sc ii} 4686{\AA} emission (Clark et al. \cite{clark00}, Hynes et al. \cite{hynes}). Indeed, He\,{\sc ii} 4686{\AA}
 only approached the strength observed in 
S18 when CI Cam exhibited  an X-ray flare in 1998 April. The outburst was  replicated at optical wavelengths, 
where 
CI Cam brightened by over 2.5 magnitudes (Clark et al. \cite{clark00}); to date no such behaviour  has been observed for 
S18.

Conversely, there are a number of observational precedents that suggest that
S18 could comprise a massive interacting binary system. Firstly, both GG Carina and VFTS698 are 
spectroscopically confirmed sgB[e] binaries with apparent  circumbinary discs/toroids 
(Gosset et al. \cite{gosset85}, Kraus et al. \cite{kraus12}, Dunstall et al. 
\cite{dunstall}), while Kastner et al. 
(\cite{kastner06}, \cite{kastner10}) suggest such  a configuration  from 
their modeling of the mid-IR excess of sgB[e] 
stars. Secondly, the X-ray luminosity of S18 
is well within the range observed for  colliding wind systems 
($\sim 10^{32}-10^{35}$ergs$^{-1}$; Gagne et al. \cite{gagne}) and is comparable to that of  the  sgB[e]
star Wd1-9; a massive compact interacting  binary within the starburst cluster Wd1 
 (Clark et al. \cite{clark08}, \cite{clark13}).

Finally, variable He\,{\sc ii}  4686{\AA} emission is a key observational signature of 
 a number of massive binary systems, where it is the result  of an additional contribution to the ionising flux 
from shocks arising in a  wind-wind collision zone. Under such a scenario we might explain  the increase in 
strength of the He\,{\sc ii} 4686{\AA} line{ \em  in tandem} with the apparent transition to a hot, WN9h phase as 
a  
result of the increase in wind velocity expected for a star of such a 
spectral type (Table 1), which in turn would lead to a stronger wind 
collision shock.

In this regard three LBVs are of interest. $\eta$ Carina is an eccentric,  long period ($P_{\rm orb} \sim 5.52$yr), colliding wind binary that 
shows orbital modulation in the He\,{\sc ii} 4686{\AA} line (Steiner \& Daminelli 
\cite{steiner}; Teodoro et al. \cite{teodoro}) - albeit much weaker that than seen in S18 - 
 and strong X-ray emission (maximum $L_{\rm X} \sim 1.3{\times}10^{35}$erg s$^{-1}$; Ishibashi et al. \cite{ishibashi}).

The second, HD~5980, is a close, eccentric, 
eclipsing binary  ($P_{\rm orb} \sim 19.26$d; Sterken \& Breysacher \cite{5980}) comprising an LBV primary (WN3$\Leftrightarrow$WN11h/B1.5 Ia$^+$;
Niemela \cite{niemela}, Koenigsberger et al. \cite{koenigsberger94}, Drissen  et al. \cite{drissen}, Heyardi-Melayari et al. \cite{HM})
and a WN4 secondary (Breysacher et al. \cite{breysacher82}) with a possible third O4-6 component in a wide orbit (Foellmi et al. \cite{foellmi}). 
Highly variable He\,{\sc  ii} 4686{\AA} emission has been observed both before and after its 1993-5  outbursts 
(e.g. Moffat et al. \cite{moffat},
Breysacher \&  Fran\c{c}ois  \cite{breysacher00}, Koenigsberger et al. \cite{koenigsberger10}). This is  likely 
the result of  wind/wind collisional effects, although an intrinsic
 contribution from one or both stellar winds  cannot be excluded. Unlike S18, this emission does not appear to be 
absent at any (orbital) phase. X-ray emission is present and of a comparable magnitude to S18 ($L_{\rm X} 
\sim 1.7{\times}10^{34}$erg s$^{-1}$; Naz\'{e} et al. \cite{naze02}, \cite{naze07}).

The final star is the SNe imposter SN 2000ch, which also demonstrates an asymmteric doubled lined He\,{\sc ii} 4686{\AA} line in quiescence (Pastorello
et al.  \cite{ch}). These authors draw direct comparison between SN 2000ch,  S18 and indeed HD~5980 on this basis, suggesting that all three
 could be binaries  and we note that both SN 2000ch and HD~5980 
also exhibit the rapid photometric variability seen in  S18, albeit of much greater amplitude. 

Therefore the multiwavelength properties of S18 are at least consistent with 
other luminous colliding wind binaries containing  LBVs.
 Given the unprecented strength and variability in the He\,{\sc ii} 4686{\AA} line it would be of considerable interest to search for periodic
 variability  indicative of orbital modulation, especially since the inclination of S18  (Zickgraf et al. 
\cite{zickgraf89}) disfavours identification of binary 
reflex motion and no evidence of  eclipses or elipsoidal modulation is present in the lightcurve. Moreover, higher S/N X-ray observations in order to 
distinguish between relativistic-accretor and colliding-wind scenarios would be of considerable interest.

\section{Concluding remarks}

Photometric and spectroscopic observations of S18 over the past half century 
 demonstrate  that it is a highly dynamic and complex system; at odds with the 
historic picture  of sgB[e] stars as essentially static objects. The cool circumstellar disc/torus 
shows little spectroscopic variability over the course of this period; appearing to be essentially decoupled
from the central stellar source, which is delineated by the high excitation
emission component of the composite spectrum and {\em is} highly variable. 

Our observations reveal that, trivially, S18 satisfies the eponymous classification criteria of LBVs. However, this does not imply that
the physical causes of its variability are the same as those of canonical S-Doradus type objects such as AG Car. Specifically, the (assumed) 
`stellar' spectrum -  and the speed of its evolution - is more extreme than canonical examples. Moreover, the  photometric variability of S18 is 
difficult to interpret in the framework 
of known LBVs, although the rapidity of the `flaring' does appear to 
mirror that of some `SNe imposters'. The colour behaviour is likewise atypical and, surprisingly, we see no 
apparent correlation between spectroscopic and  photometric behaviours, raising the possibility that the bolometric luminosity of S18 
varies over time.

The detection of X-ray emission adds to the weight of evidence that S18  hosts
 an unseen binary companion, with observations apparently favouring a colliding wind system over a relativistic accretor, although the latter 
 cannot currently be excluded. In this respect the combination of a powerful
 wind at relatively  modest  luminosity inferred by Zickgraf et al.
 (\cite{zickgraf89}) is of 
interest, since it mirrors the properties of the BHG Wd1-5, which is thought 
to have evolved through a phase of binary-driven mass-loss (Clark et al. \cite{science}).  Model atmosphere 
analysis of S18 to determine if it shares the anomalous chemistry of Wd1-5 would be welcome,  but is 
currently premature given its aspherical
 circumstellar envelope and the possibility of spectral contamination resulting 
from binarity.

These observational properties highlight the importance of S18 for a number of 
reasons. Firstly it strengthens the physical link posited 
between sgB[e] stars  and LBVs and also the possibility that the B[e] 
phenomenon is linked with binarity; specifically that the circumstellar/binary 
disc is the direct result of binary evolution (e.g. Zickgraf \cite{zickgraf03},
Kastner et al. \cite{kastner10}, Kraus et al \cite{kraus12}). The presence 
of a detached, quasi-Keplerian viscous disc/torus is attractive since the 
decadal viscous timescale of such a structure would  naturally explain the  relative lack of variability in its 
spectral signatures in comparison to the central star(s). Indeed, to the best of our knowledge, and in contrast 
to  classical Be stars, disc-loss and/or reformation has not been observed in 
sgB[e] stars. This would be explicable if the viscous 
(and hence  dissipation/formation) timescales  of discs around sgB[e] stars 
is significantly longer than that of classical Be stars due to their 
greater physical extent and lower temperatures.

Recent observational studies suggest both a high 
binary fraction amongst massive stars and that the majority of these systems
will in turn interact (Sana et al. \cite{sana}). Given this, the suggestion 
that sgB[e] stars may be currently undergoing, or have recently undertaken, 
significant  binary-driven mass-loss/transfer is important,
 since
it would allow for their identification with this brief evolutionary phase.
Other examples of this phenomenon would then be the sgB[e] star Wd1-9 (Clark et al. \cite{clark13}) and 
RY 
Scuti (e.g. Smith et al. \cite{smith11}); the latter system is of particular relevance since the authors speculate that the mass transfer in 
this system is mediated by episodic, LBV-like events. 

 Finally, the presence of a massive detached disc/torus associated with an apparent LBV raises issues regarding the nature of
its demise, given recent observational and theoretical predictions that such stars can serve as the immediate
progenitors of core-collapse SNe (e.g. Gal-Yam \& Leonard \cite{galyam}, Groh et al. \cite{groh13}). Indeed the 
latter authors find  this occurs for stars in the range log$(L_{\rm bol}/L_{\odot}) \sim 5.3-5.6$ ($M_{\rm initial}\sim 20-25M_{\odot}$); 
directly spaning the luminosity range we infer for S18 (Sects. 5 and 5.1). Following from this it 
is immediately tempting to  draw parallels between S18 and the B3 Ia progenitor of SN1987A (Walborn et al. 
\cite{walborn}), which was of similarly modest luminosity (log($L_{\rm bol}/L_{\odot})\sim5.1\pm0.1$), must
 have been  associated with a circumstellar disc and for which a binary interaction/merger scenario has been proposed
that  would have left the progenitor resembling a sgB[e] star in its immediate aftermath  
(e.g. Morris \& Podsiadlowski \cite{mandp}).

Another intriguing comparator is SN2009ip. Initially classified as a SNe-imposter, between 2009-11 it   demonstrated
 rapid photometric  variability of a similar, but more extreme nature than S18 (Sect. 5.1; Smith et al. \cite{smith10}, 
Foley et al. \cite{foley}, Pastorello et al. \cite{ip}).  During this period, its spectrum 
was also not dissimilar to that of S18 in gross terms, being dominated by Balmer series He\,{\sc i} and Fe\,{\sc ii} line emission in the blue and Paschen series, 
Ca\,{\sc ii} 
and O\,{\sc i} emission in the red, although outflow velocities were substantially 
greater than implied for  S18 and a  greater mass ($\sim 50-80 M_{\odot}$)  also inferred. In 2012 it displayed two eruptions; the nature of the first event 
is uncertain, with core-collapse SNe, pair-production instability and binary merger all being proposed, while the second is assumed to result from the 
interaction of  rapidly expanding ejecta with existing circumstellar material (Fraser et al. \cite{fraser}, Mauerhan et al. \cite{mauerhan}, Pastorello et al. \cite{ip}, Prieto et al. 
\cite{prieto}, Soker \& Kashi \cite{soker}). Both Smith et al. (\cite{smith13}) and Levesque et al. (\cite{levesque}) conclude that dust was present in the 
circumstellar material, with the latter authors suggesting a disc-like geometry and a possible origin in binary-driven mass-loss (although an origin in the recent 
photometric outbursts is also viable). As such, the possibility that S18 represents a lower mass analogue of this highly unusual energetic transient is particularly 
exciting.

\begin{acknowledgements}
We thank the anonymous referee for their insightfull comments which have greatly improved this manuscript.
This research is partially supported by the Spanish Ministerio de Ciencia e Innovaci\'on (MICINN) under 
grant AYA2010-21697-C05-01/05.
The OGLE project has received funding from the European Research Council
under the European Community's Seventh Framework Programme
(FP7/2007-2013) / ERC grant agreement no. 246678 to AU. The AAT observations have been supported by the OPTICON
project (observing proposals 2011A/014 and 2012A/015), which is funded by the European Commission under the 
Sevent Framework Programme (FP7).
Drs Carlos Gonz\'{a}lez-Fern\'{a}ndez and Amparo Marco prepared and executed some of 
the AAT observations. Finally, we wish to 
thank Paco Najarro, Atsuo Okazaki and Jose Groh for their valuable input. 

\end{acknowledgements}

{}

\appendix

\section{Summary of historical and current data for S18}

\begin{table*}
\begin{center}
\caption[]{Optical spectroscopic summary}
\begin{tabular}{lclr}
\hline
\hline
Date &I-band & Comments & Reference \\
\hline
$<1955$ & - & H$\alpha$, H$\beta$, H$\gamma$ and He\,{\sc ii} 4686{\AA} in emission & Lindsay (\cite{lindsay}) 
\\[1mm]

1967 July 14 & - &  Balmer emission to H$\epsilon$. He\,{\sc ii} 4686{\AA} absent  & Sanduleak (\cite{sanduleak}) 
\\[1mm]

1972-74 & - & H$\alpha$, H$\beta$, H$\gamma$, H$\epsilon$ in emission. No mention of He\,{\sc ii} 4686{\AA} & 
Azzopardi et al. (\cite{azzopardi75}) \\[1mm]

1977 September 17 & - & Balmer emission to H$\epsilon$. He\,{\sc ii} 4686{\AA} emission comparable to H$\beta$  & 
Sanduleak (\cite{sanduleak}) 
\\[1mm]

1978 January 27 & - & H$\beta$ to H13 P Cygni emission.  He\,{\sc ii} 4686{\AA} 
and Fe\,{\sc ii} (multiplet 24) in emission & Azzopardi et al. (\cite{azzopardi}) \\[1mm]

1978 July 16 & - & H$\alpha$ to H11 and He\,{\sc ii} 4686{\AA} in emission & Shore et al. (\cite{shore87}) \\[1mm]

1981 July 13 & - &   Balmer series and He\,{\sc ii} 4686{\AA}  in emission & Shore \& Sanduleak \cite{shore82} 
\\[1mm]

1982 March 22 & - & Balmer series and He\,{\sc ii} 4686{\AA}  in emission & Shore \& Sanduleak \cite{shore82} \\
              &  & He\,{\sc ii} 4686{\AA} $\sim$50\% weaker than 1981 July 13 &     \\[1mm]       
 
1983 Oct 10 & - &H$\alpha$ to H11 and He\,{\sc ii} 4686{\AA} in emission & Shore et al. (\cite{shore87}) \\[1mm]

1983 December 17 & - &  H$\alpha$ to H11  and He\,{\sc ii} 4686{\AA} in emission. &  Shore et al. (\cite{shore87}) 
\\      
 & &  P Cygni profile for higher Balmer lines but not H$\alpha$ or He\,{\sc ii} 4686{\AA}  &  \\[1mm]

1987 Nov. 7 & - & H$\alpha$ to H$\delta$ and He\,{\sc i} 5876{\AA} in P Cygni emission. He\,{\sc ii} 4686{\AA} 
absent
& Zickgraf et al. (\cite{zickgraf89}) \\[1mm]

1991 September 17-20& -  &H$\alpha$ to H$\delta$ and He\,{\sc i} in emission.  Double peaked He\,{\sc ii} 4686{\AA}
& Nota et al. (\cite{nota}) \\ [1mm]

2000 October 11-14 & 12.52 & H$\beta$ to H$\delta$ in emission. He\,{\sc ii} 4686{\AA} absent  & Massey \& Duffy 
(\cite{massey}) \\[1mm]

2000 October 13 & 12.52 &H$\alpha$, H$\beta$ and He\,{\sc i} 6678{\AA} in P Cygni emission. He\,{\sc ii} 4686{\AA} 
absent & Torres et al. (\cite{torres}) \\ [1mm]

2001 November 24 &12.66 & H$\alpha$, H$\beta$ and He\,{\sc i} 6678{\AA} in  emission. He\,{\sc ii} 4686{\AA} absent 
& Torres et al. 
(\cite{torres}) \\[1mm] 

2005 December 10 &12.69 & H$\alpha$, H$\beta$, He\,{\sc i} 6678{\AA} and  He\,{\sc ii} 4686{\AA} in emission & 
Torres et al. 
(\cite{torres}) \\[1mm] 

2007 October 3-4& 13.10  &H$\alpha$, H$\beta$, He\,{\sc i} 6678{\AA} and  He\,{\sc ii} 4686{\AA} in emission & 
Torres et al. 
(\cite{torres}) \\[1mm] 

2008 November 13  & 12.79 & H$\alpha$, H$\beta$, He\,{\sc i} 6678{\AA} and  He\,{\sc ii} 4686{\AA} in emission & 
Torres et al. 
(\cite{torres})  \\[1mm]

2010 November 24 & 12.55 & H$\beta$ to H$\epsilon$ in P Cygni emission, He\,{\sc ii} 4686{\AA} absent & This work 
\\[1mm]

2011 July 11 &12.62    & H$\alpha$ to H$\delta$, He\,{\sc i} 6678{\AA} and  7065{\AA} and He\,{\sc ii} 4686{\AA} in 
emission & This work \\
           &      & He\,{\sc ii} 4686{\AA} stronger than H$\gamma$  & \\[1mm]

2011 December 8 & 12.51 &H$\beta$ to H$\epsilon$ in emission, He\,{\sc ii} 4686{\AA} weakly in emission & This work 
\\
 &               & H$\beta$ single peaked, H$\gamma$ to H$\epsilon$ P Cygni profiles & \\[1mm]

2012 July 07 &12.92    & H$\beta$ to H$\epsilon$ in (single peaked) emission, He\,{\sc ii} 4686{\AA} stronger than 
H$\gamma$ & This work \\

\hline
\end{tabular}
\end{center}
I-band magnitudes quoted are from  OGLE-II,-III and -IV (Sect. 2.1). Note that emission lines are single peaked 
unless otherwise noted. Lindsay (\cite{lindsay}), 
Sanduleak (\cite{sanduleak}), Azzopardi et al. (\cite{azzopardi75}),
Shore \& Sanduleak  (\cite{shore82}). Shore et al.
(\cite{shore87}) and Torres et al. (\cite{torres}) do not present all the spectra described in these works.
 Additionally
Azzopardi \& Breysacher (\cite{azzopardi79}) detected  He\,{\sc ii} 4686{\AA} in emission in 1977 October 
via an objective prism survey. 
\end{table*}

\section{Colour magnitude behaviour of LBVs and SNe imposters}

\begin{figure*}
\includegraphics[width=8cm,height=15cm,angle=270]{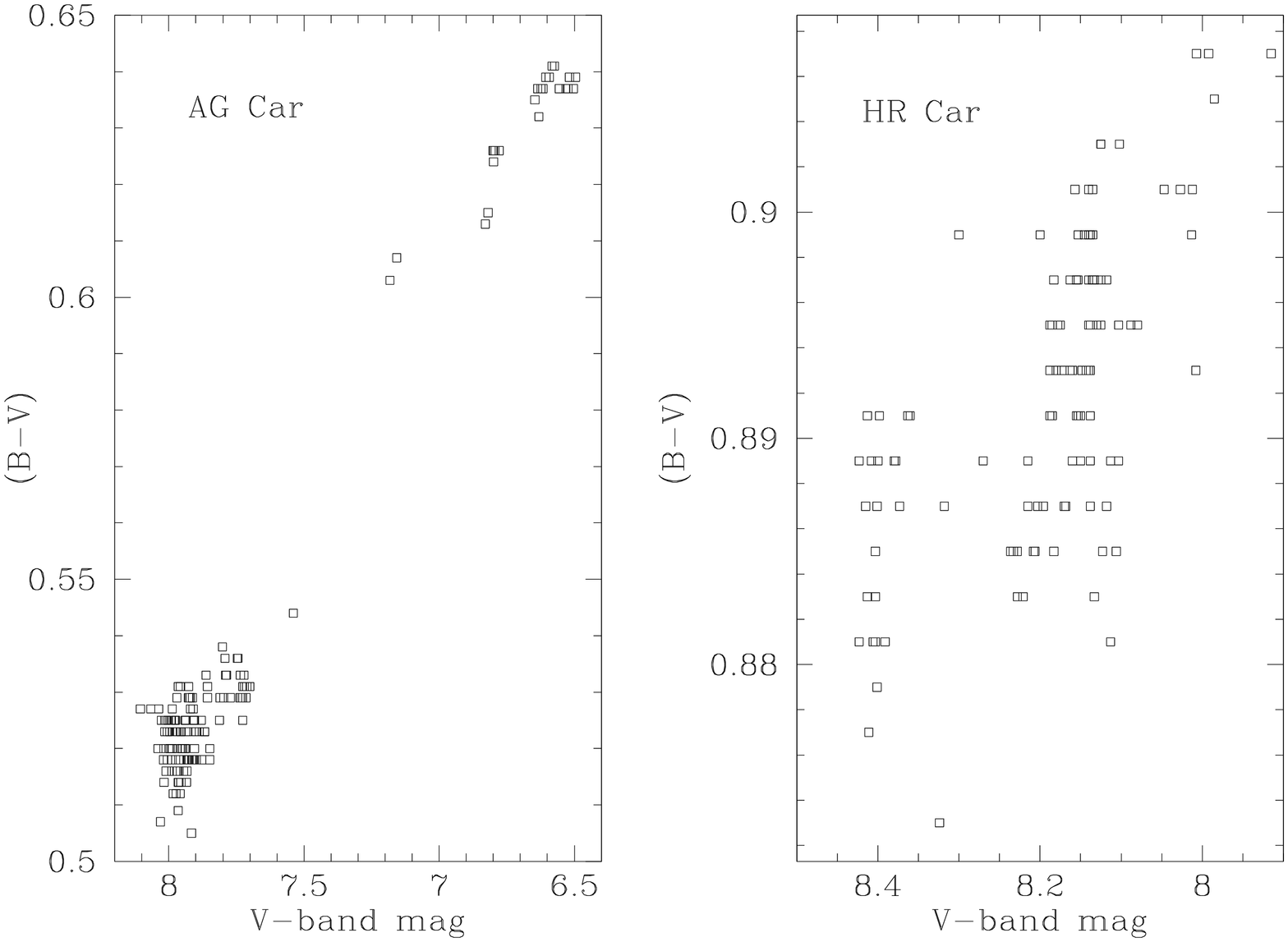}
\includegraphics[width=8cm,height=15cm,angle=270]{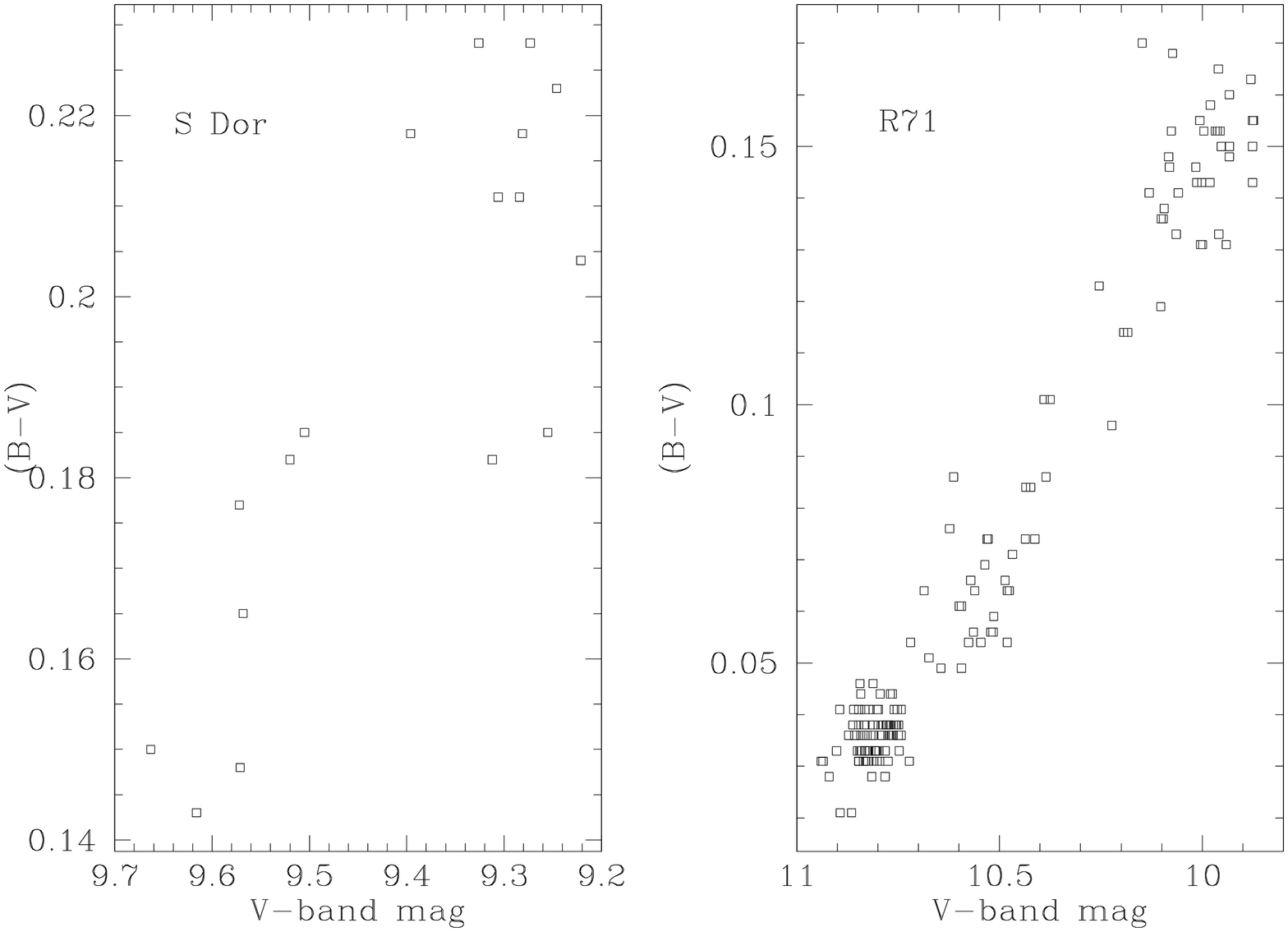}
\includegraphics[width=8cm,height=15cm,angle=270]{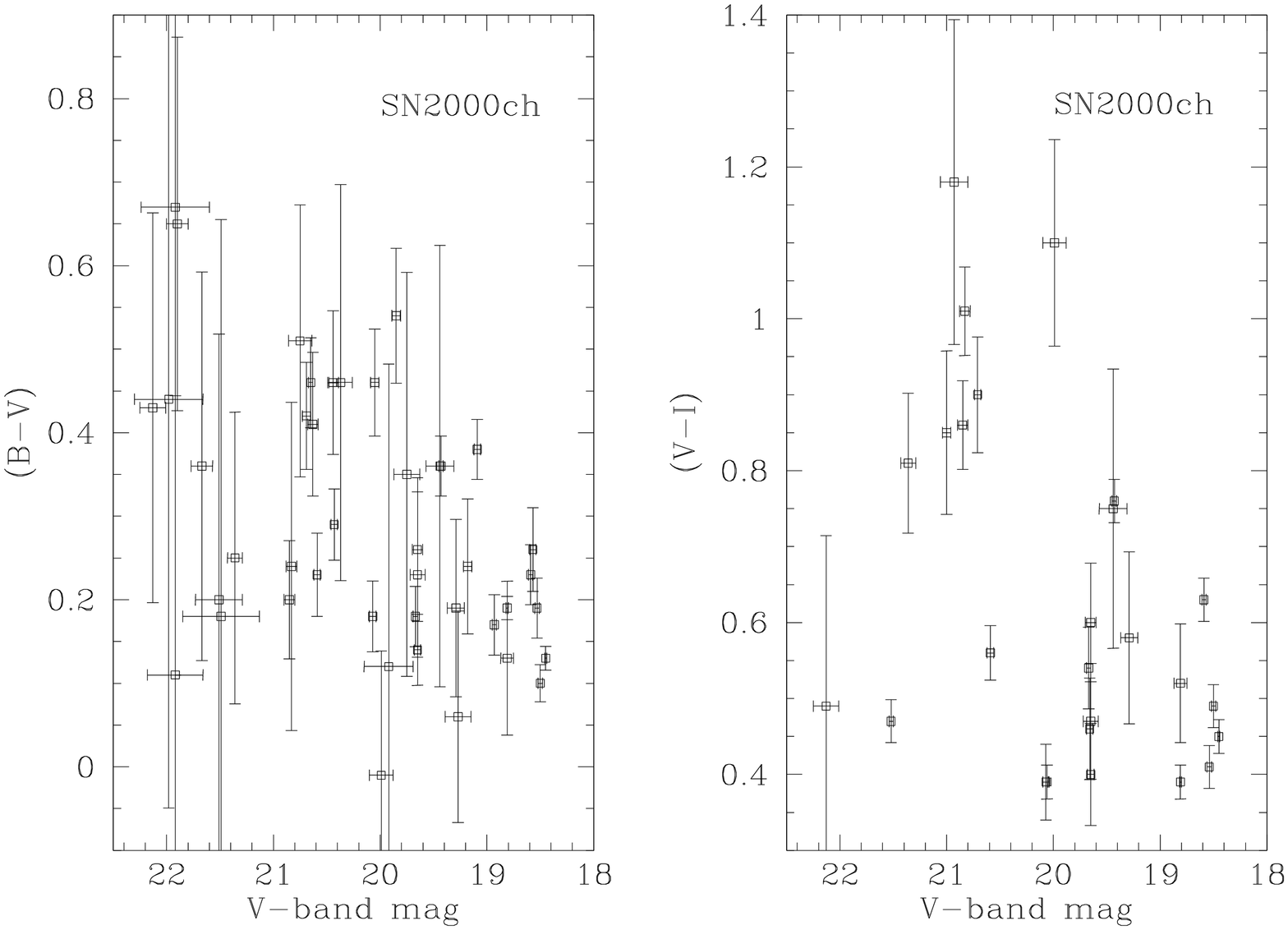}
\caption{Colour magnitude plots of 4 known LBVs in the Galaxy and Magellanic clouds and
and the SNe imposter SN2000ch (=NGC2403 LBV1).}
\end{figure*}

\end{document}